\pdfoutput=1

\documentclass[letterpaper,10pt,twocolumn, final]{article} 
\usepackage{amsmath}
\usepackage{float}
\usepackage[latin1]{inputenc}
\usepackage{graphicx}
\usepackage{amsthm}
\usepackage{amssymb}
\usepackage{amsmath}
\usepackage{cite}
\usepackage{url}
\usepackage{balance}
\usepackage{algorithm}
\usepackage{algpseudocode}
\usepackage{pifont}
\usepackage[toc,page]{appendix}

\usepackage{lipsum}
\usepackage{enumitem}
\setlist[enumerate]{itemsep=0mm}

\newtheorem{mytheorem}{Theorem}

\usepackage[top=1in, bottom=1in, left=0.625in, right=0.625in]{geometry}
\usepackage{titling}
\pretitle{\noindent\begin{center}\large\bfseries\MakeUppercase}
\posttitle{\end{center}\noindent}
\usepackage{abstract}

\renewcommand\thesection{\arabic{section}.}
\renewcommand\thesubsection{\thesection\arabic{subsection}.}
\renewcommand\thesubsubsection{\thesubsection\arabic{subsubsection}.}

\usepackage{titlesec}
\titlespacing*{\section}{0pt}{12pt}{10pt}
\titlespacing*{\subsection}{0pt}{12pt}{10pt}
\titlespacing*{\subsubsection}{0pt}{12pt}{0pt}

\titleformat{\section}{\center\normalfont\bfseries}{\thesection}{1em}{\MakeUppercase}
\titleformat{\subsection}{\normalfont\bfseries}{\thesubsection}{1em}{}
\titleformat{\subsubsection}{\normalfont\itshape}{\thesubsubsection}{1em}{}

\usepackage{tgtermes}
\usepackage[T1]{fontenc}

\usepackage[small]{caption}
\floatstyle{plaintop}
\restylefloat{table}

\makeatletter
\def\bstctlcite{\@ifnextchar[{\@bstctlcite}{\@bstctlcite[@auxout]}}
\def\@bstctlcite[#1]#2{\@bsphack
      \@for\@citeb:=#2\do{%
        \edef\@citeb{\expandafter\@firstofone\@citeb}%
        \if@filesw\immediate\write\csname #1\endcsname{\string\citation{\@citeb}}\fi}%
      \@esphack}
\makeatother

\usepackage{authblk}

\date{\vspace{-2em}}

\begin{document}

\title{Wideband Distributed Spectrum Sharing with Multichannel Immediate Multiple Access}
\author{Mingming Cai (mcai@nd.edu)\thanks{This paper is an extended version of the work published in \cite{mingmingcaiJNLWInnComm2016}. }}
\author{J. Nicholas Laneman (jnl@nd.edu)}
\affil{University of Notre Dame, Notre Dame, IN, 46556, USA}
\renewcommand\footnotemark{}
\renewcommand\footnoterule{}

\maketitle

\bstctlcite{biblio:BSTcontrol}

\begin{abstract}
This paper describes a radio architecture for distributed spectrum sharing of multiple channels among secondary users (SUs) in a wide band of frequencies and a localized area. A novel Multichannel Immediate Multiple Access (MIMA) physical layer is developed such that each SU can monitor all the channels simultaneously for incoming signals and achieve fast rendezvous within the multiple channels.
The spectrum utilized by an SU pair can be changed dynamically based upon spectrum sensing at the transmitter and tracking synchronization and control messages at the receiver. Although information about the number of active SUs can be used to improve the spectrum sharing efficiency, the improvement is small relative to the cost of obtaining such information. Therefore, the architecture adopts Multichannel Carrier Sense Multiple Access (CSMA) for medium access control regardless of the number of active SUs. A prototype implementation of the architecture has been developed using an advanced software defined radio (SDR) platform. System tests demonstrate that the spectrum sharing efficiency of the prototype is close to an upper bound if the signal-to-noise ratio (SNR) is sufficiently high. Among other practical issues, imaged interference caused by hardware IQ imbalance limits system performance. In the prototype, the MIMA is based on an LTE waveform. Therefore, the spectrum sharing radio can be potentially applied to the 3.5~GHz radar band for Citizens Broadband Radio Service (CBRS).
\end{abstract}

\section{Introduction}
\label{sec:introduction}
In 2015, global mobile data traffic grew by 74 percent while mobile network connection speeds only grew by 20 percent. Mobile traffic is predicted to continue growing at 53 percent annually over the next five years \cite{cisco2016}. The astonishing growth of wireless devices and applications have led to exploding demands for radio spectrum. In addition, government reports suggest that enhancing access to spectrum plays an important role in both economic growth and technological leadership, e.g,. \cite{PCASTSpectrumReport2012}.  Among other initiatives, the DoD National Spectrum Consortium \cite{NSC2016} and the NSF Advanced Wireless Initiative \cite{NSF2016} were created to accelerate industry, academic, and government efforts to develop technologies and policies for better spectrum utilization in the United States (US).

There are two basic directions to exploit more spectrum resources: sharing current under-utilized bands through cognitive radio (CR) and dynamic spectrum access (DSA), and exploring spectrum opportunities in higher frequencies, e.g., millimeter-wave bands \cite{mingmingcaiJNLWInnComm2016, cui2014unified}. Millimeter wave signals experience much higher path loss than signals below 6~GHz \cite{el2014spatially}, and millimeter wave systems require sophisticated array beamforming architectures to overcome this path loss. Therefore, sharing under-utilized spectrum has advantages such as broader coverage and lower cost.

Three major blocks of radio spectrum are currently being explored for sharing in the US: television white space (TVWS), 3.5~GHz radar band for Citizens Broadband Radio Service (CBRS), and Unlicensed National Information Infrastructure (U-NII) spectrum at 5~GHz. TVWS are broadcast TV channels at 470-790~MHz in Europe and discontinuous 54-698~MHz in the United States \cite{ESTITVWS2014, adrianabfloresryaneguerraedwardwknightly2013}. In 2015, the FCC established the CBRS for shared wireless broadband use of the 3550-3700~MHz band (3.5~GHz Band) \cite{FCCReportOrder2015}. The National Telecommunications and Information Administration (NTIA) is working on sharing 195~MHz of spectrum in the 5~GHz band with U-NII devices \cite{NTIAReport2015}, specifically, 5350-5470~MHz and 5850-5925~MHz. The three blocks of spectrum can be shared in time, frequency and space \cite{QingZhaoandBMSadler2007, ZSunandJNlaneman2014}.

The President's Council of Advisors on Science and Technology (PCAST) and the FCC have recommended that the shared spectrum be classified into three tiers \cite{PCASTSpectrumReport2012, FCCReportOrder2015}. The first tier is the legacy or incumbent users, which would be granted full protections for operations within deployed areas. The second tier consists of users with a short-term Priority Access (PA) license allowing them to operate in designated spectrum and geographic areas. The second-tier users receive protection from interference of the third-tier users, but are subject to the interference from the first-tier users. The third-tier users are General Authorized Access (GAA) operators, which could only utilize the spectrum on an opportunistic basis, and no interference protection is provided. In this paper, the first- and second-tier users correspond to primary users (PUs), and the third-tier users correspond to secondary users (SUs) in the traditional DSA context.
In DSA, PUs have legacy rights or high priority to use the spectrum, and SUs can opportunistically access the spectrum without causing harmful interference to the PUs \cite{GoldsmithANDJafarANDMaric2009}.

SUs can dynamically operate in wideband spectrum to identify more spectrum opportunities. There are two key issues that need to be addressed for wideband DSA: identification of spectrum opportunities and assignment of spectrum resources for SUs \cite{mingmingcaiJNLAllerton2015}.

Databases and spectrum sensing are key approaches for identification of spectrum opportunities. A database can assist identification of spectrum opportunities by directly providing PUs' spectrum usage information and by improving the quality of spectrum sensing with detailed signaling parameters and prior information, such as PU power levels, locations and dwell times \cite{franciscopaisanajoaopaulomirandanicolamarchettluizadasilva2014}. Energy detection, matched filter detection, and feature detection are typical methods for single-band spectrum sensing \cite{HanoWangGosanNohDongkyuKimSungtaeKimandDaesikHong2010}. Wideband spectrum sensing algorithms have also proposed to allow opportunistic access in wideband spectrum \cite{ZSunandJNlaneman2014}.

In a multichannel system, link establishment requires two nodes to \emph{rendezvous}, that is, two radio nodes find one another in one of the multiple available channels \cite{NickTheisRyanThomasLuizDasilva2001}. There has been significant research on medium access control (MAC) algorithms for improving the performance of spectrum sharing systems and wireless ad hoc networks. A large number of them focus on the spectrum sharing of a single channel, such as Carrier Sense Multiple Access with collision avoidance (CSMA/CA) and CSMA with collision detection (CSMA/CD). Most of the MAC algorithms for multiple channels assume the receiver can only listen to one of the channels at a time. Two types of solutions have been proposed to allow the receiver to track the channel of the transmitter:
\begin{itemize}
  \item Channel hopping: two nodes or multiple nodes hop frequencies among available channels until they rendezvous in the same channel and establish a communication link \cite{PBahlandRChandraandJDunagan2004, kaiguibianjungminpark2011, rohangandhichihwangcharliehu2012}. The transmitter and receiver have their own channel sequences defined by the protocols. Two nodes usually cannot rendezvous in one or perhaps a few slots, even without interference from other radios in the system. Rendezvous would be much longer for multiple SU pairs considering the interference among different pairs of radios.
  \item Common control channel: a common control channel is used for the transmitter and receiver to share control information \cite{shihlinwuchunglichihyulinyucheetsengjangpingsheu2000, so2004multi,liangpingmaxiaofenghanchienchungshen2005}. The transmitter can tell the receiver its operating channel through a control message. However, if the control channel can be preempted by a PU, then the secondary system will likely fail.
 \end{itemize}
There are also MAC protocols assuming the receiver can observe all the channels simultaneously, e.g.,\cite{choi2006multichannel} considers a slotted orthogonal frequency-division multiple access (OFDMA) uplink and discusses a fast retransmission algorithm if collisions occur in multichannel random access. However, the packet size is fixed and has to be transmitted in one time slot. Furthermore, \cite{nasipuri1999multichannel} develops a multichannel CSMA MAC protocol for multihop wireless systems. The algorithm is not designed for DSA scenarios, and the number of SUs is not used to improve the system performance.

In this paper, our contributions are threefold. First, we develop a radio architecture for distributed spectrum sharing. The architecture includes a database, wideband spectrum sensing, a fast-rendezvous physical layer, and a MAC algorithm. We summarize the requirements of the four system components. Based on the requirements, a fast-rendezvous physical layer, Multichannel Immediate Multiple Access (MIMA), is designed to support distributed spectrum sharing \cite{mingmingcaiJNLAsilomar2015}. Second, we develop MAC algorithms to maximize the spectrum usage. The features of the MAC algorithm include distributed spectrum access, fast rendezvous, utilization of the number of SUs to assist in spectrum sharing, and variable packet size. Analytical and simulation results both demonstrate that information about the number of SUs can increase the performance of spectrum sharing, but the improvement in performance is relatively small compared to the cost of acquiring such information. Third, we develop a wideband spectrum analyzer, and integrate it together with the MIMA physical layer and multichannel CSMA into a complete spectrum sharing system. The whole system has been developed with National Instruments (NI) USRP RIOs as well as the LabVIEW Communications System Design Suite (CSDS) and LTE Application Framework \cite{NICSDS2015}. The LTE-based prototype of the spectrum sharing radio can potentially be used for the CBRS in 3.5~GHz radar band.

The paper is organized as follows. Section~\ref{sec:system-architecture} describes the system architecture and formulates the problem of MAC algorithms with different types of SU information. The slot-level MAC algorithm with information about the number of SUs is analyzed in Section~\ref{sec:MAC-Layer-Analysis-with-Full-SU-Information}. Theoretical analysis indicates the number of SU can be used to improve system throughput. In Section~\ref{sec:MAC-Algorithms-Design}, specific MAC algorithms are designed and simulated based on the slot-level analysis. Section~\ref{sec:SDR-Prototyping-of-Multichannel-CSMA} overviews software defined radio (SDR) prototyping of the architecture with Multichannel CSMA. Section~\ref{sec:Conclusion-ongoing-work} concludes the paper with a brief summary of ongoing work.

\section{System Architecture}
\label{sec:system-architecture}

\begin{figure}
\centering
\includegraphics[width=85mm]{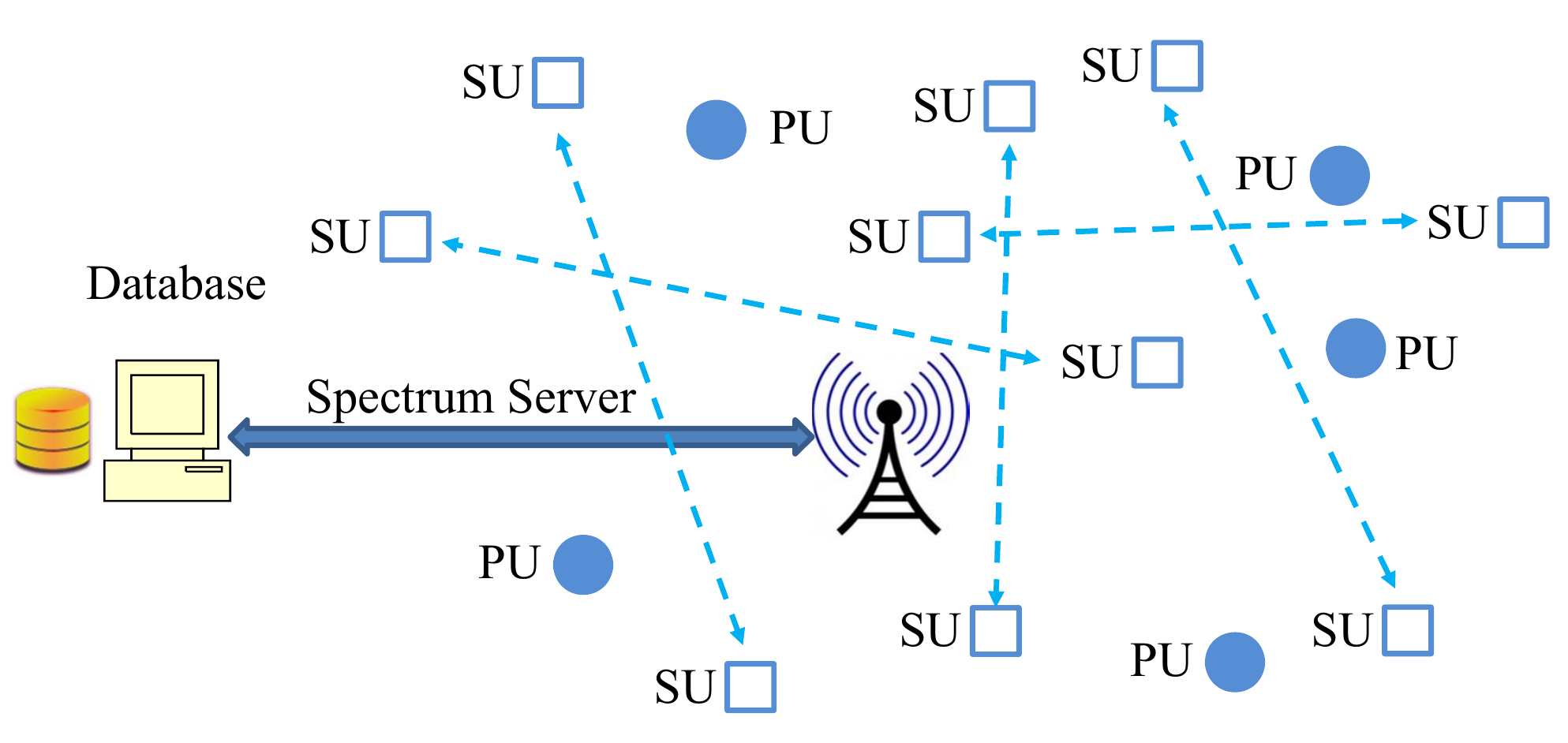}
\caption{Spectrum sharing scenario under consideration.}
\label{senario-considered}
\end{figure}

We consider a local area containing PUs and SUs, and candidates of shared spectrum as shown in Figure~\ref{senario-considered}. An SU pair has a transmitter and a receiver. A spectrum access system (SAS), consisting of spectrum server and database, protects PUs from interference by SUs. The SAS can either eliminate certain frequency and time slots from SUs' consideration or provide potential frequency and time slots available to SUs through periodic broadcast messages. Each SU uses distributed wideband spectrum sensing and a MAC protocol to share the available spectrum.

Our goal is to develop an elaborate and evolvable architecture for distributed system sharing that enables experimental validation over the air with dozens of nodes, and to develop MAC algorithms to maximize the spectrum utilization.

For the MAC analysis, we consider a slotted system so that the time slots are aligned for all the channels and transmission is restricted in slots \cite{choi2006multichannel}. A data packet can be transmitted in one or multiple slots, depending on the SUs' applications. If the slot length is much smaller than the average transmission time of a packet, the performance of the slotted system will be close to that of an unslotted system.


For the local-area spectrum sharing system, each channel is modeled as a collision channel \cite{massey1985collision}. In a collision channel, a transmission gets through if and only if there is exactly one transmitter using a channel in a slot; if more than one user transmits over a channel in the same slot, a collision occurs and no information can get through the channel \cite{mingmingcaiJNLAllerton2015}. Since slots are short, feedback to indicate collision is usually not available immediately after a slot of transmission in practice, so we assume there is no slot-level collision feedback. 

Suppose there are a total of $N_c$ channels and $M$ SUs. $N$ out of the $N_c$ channels are not used by the PUs and are available for the $M$ SUs.
Each SU can access at most one channel at a time.
At the beginning of slot $k$, suppose there are $N_k$ available channels and $M_k$ SUs trying to find channels for spectrum access. Label the accessing SUs as $1,2,\dots,M_k$. Label the available channels as $1,2,\dots,N_k$.

\begin{figure}
\centering
\includegraphics[width=85mm]{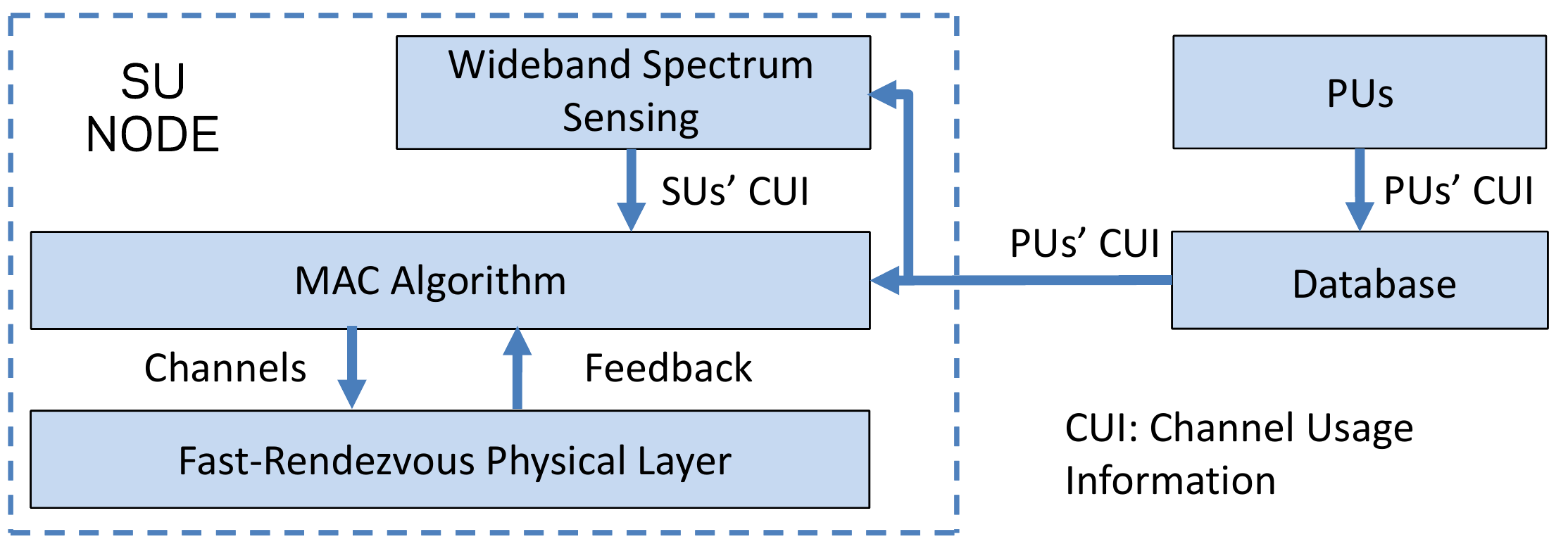}
\caption{Architecture of the wideband spectrum sharing system. The database provides PUs' channel usage information (CUI), and wideband spectrum sensing provides SUs' CUI. The feedback is packet-level feedback but not slot-level, collision feedback.}
\label{system-architecture}
\end{figure}

As shown in Figure~\ref{system-architecture}, our system architecture for wideband spectrum sharing has four components, i.e., a database, wideband spectrum sensing, fast-rendezvous physical layer, and MAC algorithm. The database and wideband spectrum sensing assist in identification of channel opportunities. Specifically, the database provides PUs' channel usage information (CUI), and wideband spectrum sensing helps to identify the channel usage of other SUs. The fast-rendezvous physical layer should be designed to meet the requirements of the MAC algorithm.  We now summarize several key features of each of these components.

\subsection{Database}

\begin{figure}
\centering
\includegraphics[width=85mm]{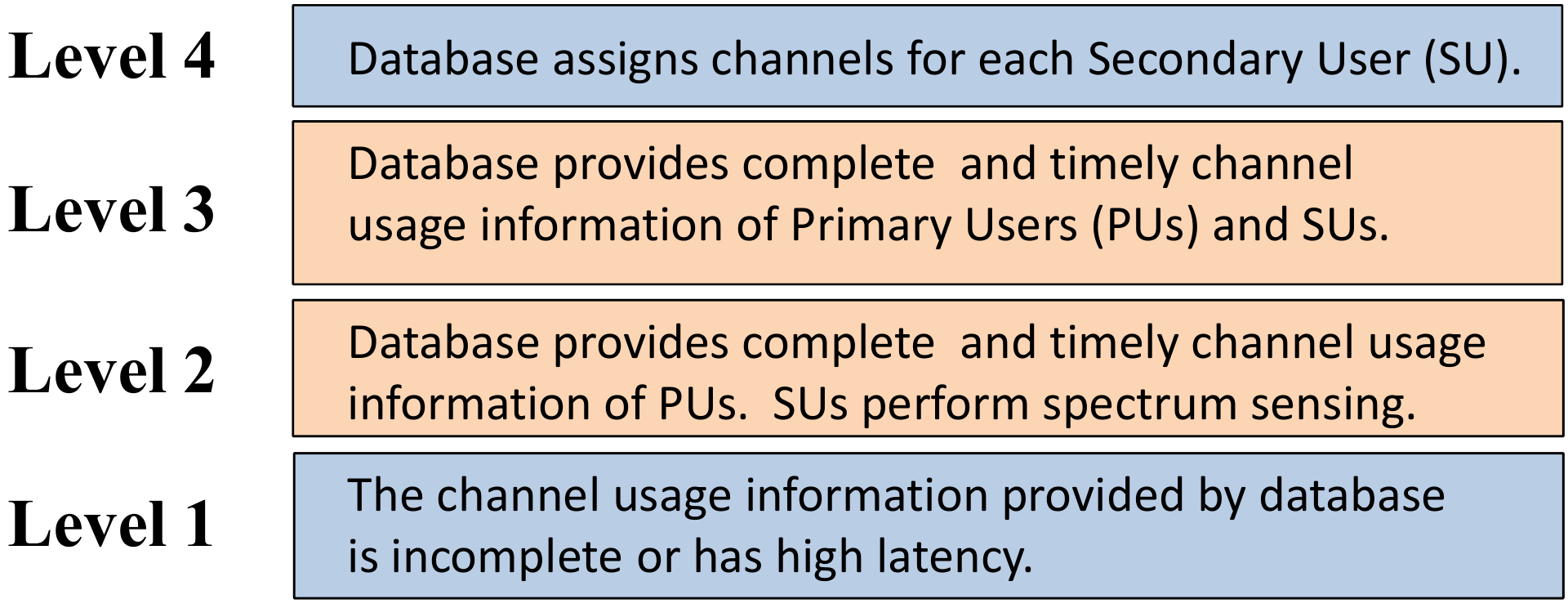}
\caption{Database levels.}
\label{fig:database-level}
\end{figure}

As discussed in \cite{mingmingcaiJNLAllerton2015}, a database can provide four levels of information. Figure~\ref{fig:database-level} illustrates the four levels. As the database level increases, the database provides more information to the SUs, but the requirements and overhead for the database also increase.

%
PUs usually have more stringent interference requirements than SUs, and SUs need timely CUI of both PUs and other SUs. We focus on the second-level database to provide good protection for the PUs. SUs can obtain correct and timely CUI for PUs from the database, and they can conduct wideband spectrum sensing to obtain CUI for other SUs. This approach seems most appropriate for scenarios in which PUs are sparse and SUs are dense.

Two communication links are required, namely, a database-SUs link and a database-PUs link. Generally, one-way mode, broadcasting PUs CUI to the SUs, is sufficient for the database-SUs link. The Database-PUs link could be either one-way or two-way, depending on the link reliability requirements.

\subsection{Wideband Spectrum Sensing}
Spectrum sensing provides the set of available channels in the sensing slot.  Wideband spectrum sensing algorithms include Nyquist sampling and sub-Nyquist sampling \cite{ZSunandJNlaneman2014}.
Fast Fourier transform (FFT)-based spectrum sensing could be applied if the hardware's sampling rate is high enough to cover the entire spectrum \cite{cai2014design}.
Sub-Nyquist sampling methods could lead to missed spectrum opportunities if there is a large number of SUs. A wideband spectrum sensing algorithm must be fast, and accurate, and it must have low probabilities of false alarm and miss detection.

We design, implement, and test an FFT-based spectrum analyzer with energy detection to execute spectrum sensing for SUs' CUI. In a local scenario, the signal-to-noise ratio (SNR) from other radios is high, and small miss detection false alarm probabilities can be achieved. For simplicity, we assume the spectrum sensing is perfect during the analysis of the MAC algorithms, i.e., there are no false alarms or missed detections.  For the SDR implementation, imperfect spectrum sensing is considered.

\subsection{Fast-Rendezvous Physical Layer with MIMA}
\label{subsec:physical-layer-requirement}

Two MIMA architectures are outlined to achieve fast rendezvous based upon orthogonal frequency-division multiplexing (OFDM) in \cite{mingmingcaiVTC2016}, and a radio prototype with MIMA is implemented with an advanced SDR platform to validate its feasibility in \cite{mingmingcaiJNLAsilomar2015}. The developed fast-rendezvous MIMAs have the following features:
\begin{enumerate}
    \item \textbf{Frequency Agile} - Transmitters can utilize any subset of the $N$ available channels. Multiple transmitters can share channels through the MAC protocol.

    \item \textbf{Fast Rendezvous} -  Each receiver monitors all channels for a sync signal of its corresponding transmitter, and passes the transmitter's control information and payload on the rendezvous channel for further demodulation and decoding. Other transmitters' signals are ignored by the receiver.

    \item \textbf{Standard-Relevant Waveforms} - Basing the architectures on standards-revelant waveforms brings the technology closer to real-world validation and adoption.
\end{enumerate}
Even though the transmitter has the capacity of accessing multiple channels simultaneously, the transmitter is still assumed to access at most one channel at a time in this paper. The MAC using multiple channels simultaneously will be discussed in the future.

We adopt NI USRP RIO 2953R hardware and LabVIEW CSDS software tools for physical layer prototyping. The powerful DSP-focused Xilinx Kintex-7 FPGA in the USRP RIO supports computationally intensive operations that are not efficient in a general-purpose processor.

NI also provides the LTE Application Framework 1.0 to implement a 3GPP-LTE Release 10 compliant Time Division Duplex (TDD) downlink transmitter and receiver \cite{NILTEwhitepapaer2015}. The LTE Application Framework is a good candidate for physical layer prototyping of our MIMA architectures because of the flexible time-frequency Physical Resource Block (PRB). In our MIMA prototype, 100 PRBs are evenly divided into ten channels. The receiver can listen to all the ten channels simultaneously. The fast-rendezvous physical layer has been implemented and validated in \cite{mingmingcaiJNLAsilomar2015}. Tests in \cite{mingmingcaiJNLAsilomar2015} demonstrates that the modified Application Framework is fairly robust to a nearby interferer.

\subsection{MAC Algorithm}

In this paper, we focus on the design of a MAC algorithm with a fast-rendezvous MIMA physical layer. Information about the number of SUs could improve the performance of the MAC algorithm. We will explore MAC algorithms with two different types of information about the number of SUs, and compare them to the case without such information.


\subsubsection{Information about the Number of SUs}
First, an SU can know $M_k$, the total number of SUs that are trying to access the available channels at the beginning of slot $k$.
$M_k$ could be obtained from the third-level database. Each SU reports its status to the database, and the database calculates $M_k$ and broadcast it to all the SUs. $M_k$ is referred to as full SU information. Although we focus on the second-level database, we assume $M_k$ is known for MAC analysis.


Second, an SU can know one bit of information that indicates either $M_k<N_k$ or $M_k \ge N_k$, which can be acquired through spectrum sensing during transmission at an SU transmitter. If a transceiver has separate transmit and receive RF circuits and antennas, it may sense the spectrum while transmitting and determine whether $M_k \ge N_k$ or $M_k < N_k$ based on past observations about the number of available channels. For example, if there is always at least one available channel in the past $N_s$ slots, the SU can infer that $M_k < N_k$; otherwise, $M_k \ge N_k$. We leave the development of such algorithms for future work. In this paper, we assume that the transmitter can learn the one-bit information of the number of SUs, which we refer to as partial SU information.

To summarize,  we consider following three types of information about SU activity:
\begin{itemize}[itemsep=-1mm]
  \item \textbf{Full SU information}: $M_k$
  \item \textbf{Partial SU information}: whether $M_k \ge N_k$ or $M_k < N_k$
  \item \textbf{No SU information}: $M_k$ is not known
\end{itemize}

\subsubsection{Performance of MAC Algorithms}
\label{sec:spectrum-formulation}
For analysis of MAC algorithms, we assume the transmitter conducts wideband spectrum sensing before each packet for one slot. At time $t$, let $D_m\left(t\right)$ be the amount of data in bits successfully received by SU $m$, and let $R_m$ be the data transmission rate of SU $m$ in bits/sec. The average data rate for SU $m$ is
\begin{align}
{G_m} = \mathop {\lim }\limits_{t \to \infty } \frac{1}{t} {D_m}\left( t \right), \ m=1,2,\dots,M.
\end{align}
The total average data rate for $M$ users is
\begin{align}
G = \sum\limits_{m = 1}^M {{G_m}}  = \mathop {\lim }\limits_{t \to \infty } \frac{1}{t} \sum\limits_{m = 1}^M {{D_m(t)}}.
\end{align}

The performance metric we adopt is spectrum sharing efficiency defined as
\begin{align}
E = \frac{G}{{\sum\limits_{m = 1}^M {{R_m}} }},
\end{align}
which is a ratio of the total average data rate to the total data rate possible without contention.


%
%
%
%

We will design and analyze  MAC algorithms with the three types of SU information, i.e., full SU information, partial SU information and no SU information. Their spectrum sharing efficiencies are denoted as $E_f$, $E_p$, and $E_n$, respectively. We will compare the MAC algorithms and select the most appropriate one for our spectrum sharing radio prototyping.

\section{Slot-Level MAC Analysis with Full SU Information}
\label{sec:MAC-Layer-Analysis-with-Full-SU-Information}
The analysis of MAC algorithms begins with a slot-level analysis using full SU information.

The receiver of an SU always monitors all the channels. The transmitter has five states:
\begin{itemize}[itemsep=-1mm]
  \item \textbf{Inactive}: SU is out of the spectrum sharing system
  \item \textbf{Monitoring}: SU monitors whether there is a packet available for transmission
  \item \textbf{Initial}: SU begins to access the spectrum and tries to find an available channel; it starts with wideband spectrum sensing
  \item \textbf{Transmitting}: SU transmits a packet on a channel
  \item \textbf{Re-rendezvous}: SU tries to access the previously used channel again, and starts with wideband spectrum sensing
\end{itemize}

\begin{figure}
\centering
\includegraphics[width=85mm]{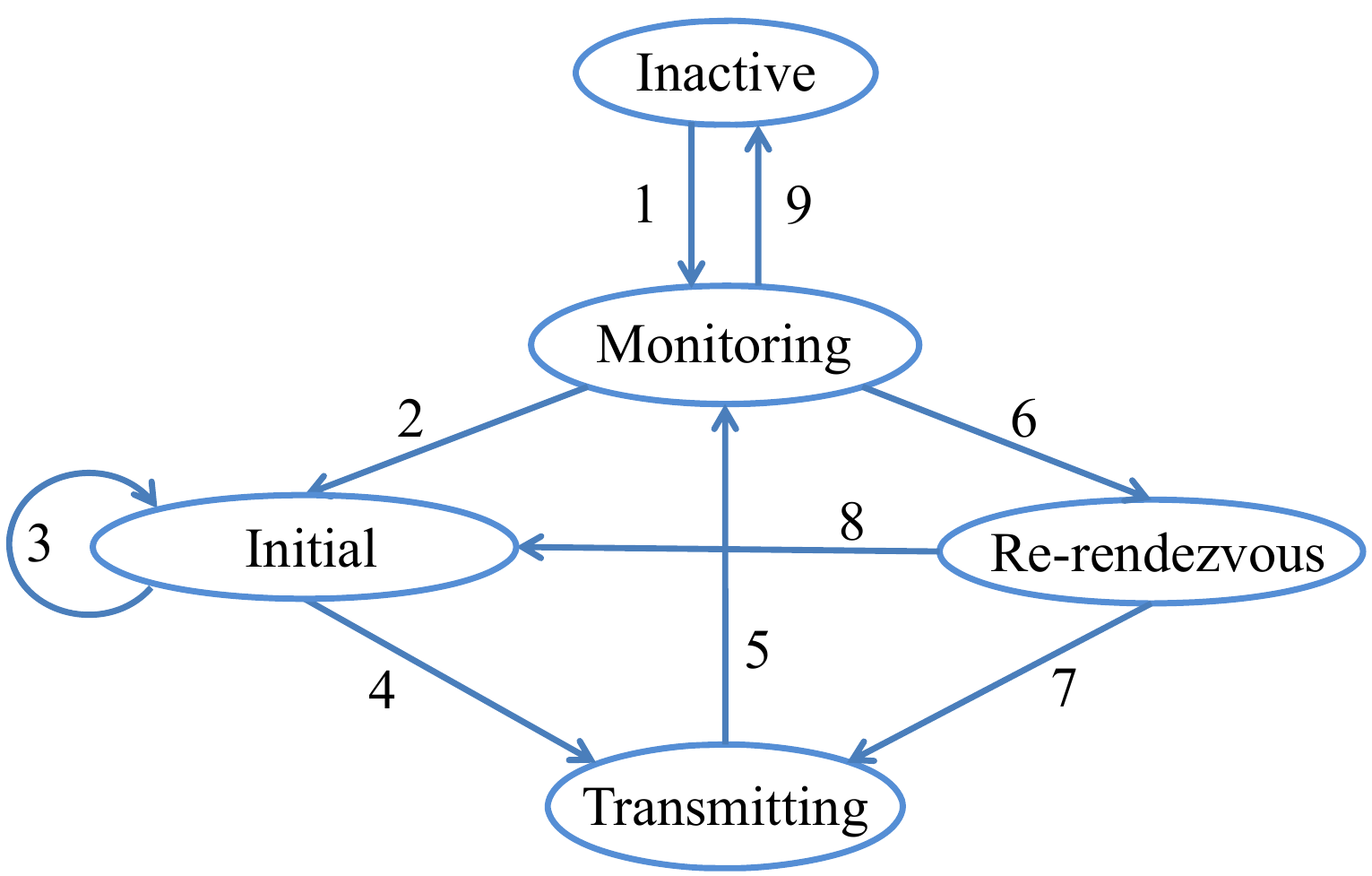}
\caption{SU Transmitter State Transition Diagram.}
\label{fig:state-transition}
\end{figure}

As shown in Figure~\ref{fig:state-transition}, there could be 12 transitions among the five states for an SU. Specially,
\begin{enumerate}[itemsep=-1mm]
  \item The SU joins the spectrum sharing system
  \item The SU has a packet for transmission but does not meet the requirement of re-rendezvous
  \item The SU does not find a channel for packet transmission, and backoff is executed
  \item The SU finds a channel to transmit a data packet
  \item The SU completes a packet transmission
  \item The SU has a packet for transmission and meets the requirements of re-rendezvous
  \item The SU finds a channel to transmit a data packet
  \item The SU does not find a channel to transmit data packet
  \item The SU leaves the spectrum sharing system
\end{enumerate}

Again, there are $N_k$ available channels and $M_k$ SUs in initial or re-rendezvous state at the beginning of slot $k$. All the $M_k$ SUs know from the database the $N_k$ available channels unused by PUs. Among the $M_k$ SUs, let $p_{m,n}$ be the probability of the $m$th SU selecting the $n$th available channel . The access distribution vector for SU $m$ in slot $k$ can be expressed as
\begin{align}
\mathbf{p}_{k,m}= \left[p_{m,1}, p_{m,2}, \dots, p_{m,N_k} \right]^T,
\end{align}
where ${\left( \cdot \right)} ^T$ denotes the transpose operation and
\begin{align}
\sum\limits_{n = 1}^{{N_k}} {{p_{m,n}}}  \le 1, \quad m=1,2,\dots,M_k.
\end{align}
If $\sum\limits_{n = 1}^{{N_k}} {{p_{m,n}}}  <1$, SU $m$ does not transmit on any channel with probability $1-\sum\limits_{n = 1}^{{N_k}} {{p_{m,n}}}$ in slot $k$.


Let $X_n=1$ be the event that there is exactly one SU pair transmitting on available channel $n$, and $X_n=0$ be the event that no SU pair or more than one SU pair transmits in the $n$th available channel. The expected number of successful transmissions without collision in channel $n$ for slot $k$ is
\begin{align}
{\mathbb{E}}\left[ X_n \right]
= \Pr \left[ {{X_n} = 1} \right]
=\sum\limits_{m = 1}^{{M_k}} {\left[ {{p_{m,n}}\prod \limits_{j = 1,j \ne m}^{{M_k}} {\left( {1 - {p_{j,n}}} \right)} } \right]},
\end{align}
where $ n=1,2,\dots,N_k$. Let $Y$ be the total expected number of successful transmissions without collision among all the $N_k$ available channels in slot $k$. Then
\begin{align}
Y\left( \mathbf{P}_k \right)
&={\mathbb{E}}\left[ \sum\limits_{n = 1}^{{N_k}} X_n \right]
= \sum\limits_{n = 1}^{{N_k}} {\mathbb{E}}\left[  X_n \right] \nonumber \\
&= \sum\limits_{n = 1}^{{N_k}} \sum\limits_{m = 1}^{{M_k}} {\left[ {{p_{m,n}}\prod \limits_{j = 1,j \ne m}^{{M_k}} {\left( {1 - {p_{j,n}}} \right)} } \right]},
\label{eq:total-throughput-in-a-slot}
\end{align}
where $\mathbf{P}_k$ is defined as
\begin{align}
\mathbf{P}_k= \left[\mathbf{p}_{k,1}, \mathbf{p}_{k,2} , \dots,  \mathbf{p}_{k,M_k} \right].
\end{align}


The goal of the MAC algorithm in the slot of channel selection becomes an optimization problem,
\begin{align}
\mathbf{P}_k^* = \mathop {\arg \max }\limits_{\mathbf{P}_k} Y\left( \mathbf{P}_k \right).
\end{align}
We will study solutions to this problem in in two scenarios, namely, all SUs in initial state, and SUs in re-rendezvous state.

\subsection{All SUs in Initial State}
For distributed spectrum access, initial-state SUs do not know the channels that the other initial-state SUs are trying to access. Therefore, the channel selection random variables should be independent for each SU. Assume all SUs are rational and make channel selection decisions on their own. All that rational SUs will have the same probability to access every channel. For example, if there is a channel with smaller collision probability, and this information is known to all SUs, all SUs will put higher access probability to access this channel. Therefore, the channel selection random variables for all SUs are identically independent distributed (i.i.d), that is,
\begin{align}
\mathbf{p}_{k}=\mathbf{p}_{k,1}= \mathbf{p}_{k,2} = \dots=  \mathbf{p}_{k,M_k} =\left[p_{1}, p_{2}, \dots, p_{N_k} \right]^T,
\end{align}
where $\sum\limits_{n = 1}^{{N_k}} {{p_n}}  \le 1$. In this case, (\ref{eq:total-throughput-in-a-slot}) can be simplified to
\begin{align}
Y\left( \mathbf{p}_{k} \right) = \sum\limits_{n = 1}^{{N_k}} M_k p_n \left( 1- p_n\right)^{M_k-1}.
\label{eq:total-throughput-in-a-slot-simplified}
\end{align}

\begin{mytheorem}
\label{theorem:max-throughput-initial}
$Y\left( \mathbf{p}_k \right)$ in (\ref{eq:total-throughput-in-a-slot-simplified}) is maximized if
\begin{align}
p_{1}= p_{2}= \dots=p_{N_k}= \min \left\{ {\frac{1}{{{N_k}}},\frac{1}{{{M_k}}}} \right\}.
\label{eq:access-probability-initial-stage}
\end{align}
The corresponding maximum is
\begin{align}
\max Y\left( {{{\bf{p}}_k}} \right) = \left\{ {\begin{array}{*{20}{c}}
   {{N_k}{{\left( {1 - \frac{1}{{{M_k}}}} \right)}^{{M_k} - 1}},\quad {M_K} > {N_k}}  \\
   {{M_k}{{\left( {1 - \frac{1}{{{N_k}}}} \right)}^{{M_k} - 1}},\quad {M_K} \le {N_k}}  \\
\end{array}} \right..
\label{eq:total-throughput-in-a-slot-simplified-max}
\end{align}

\end{mytheorem}
The theorem is proved in Appendix~\ref{appendix:proof-theorem-max-throughput-initial}. Theorem~\ref{theorem:max-throughput-initial} shows that the best strategy for an initial-state SU is to have equal probability to transmit in every available channel. The probability depends on the relation between $M_k$ and $N_k$ as shown in (\ref{eq:access-probability-initial-stage}).

\subsection{SUs in Re-rendezvous State}
Once an SU completes transmission of a packet, it enters monitoring state and waits for another packet. With a new packet, the SU tries to re-rendezvous at the beginning of slot $k$. If the previously used channel is not available, the SU switches to the initial state; if the previously used channel is available, the SU needs to decide whether to use the previously utilized channel or to switch to the initial state for rendezvous again. The discussion falls into two situations, $M_k>N_k$ and $M_k \le N_k$.

If $M_k>N_k$, some SUs have no channel for transmission. To be fair to these SUs, the re-rendezvous SUs should not access their previously used channels. Instead, all re-rendezvous SUs need to convert to the initial state and access the channel according to (\ref{eq:access-probability-initial-stage}).

If $M_k \le N_k$, we will show that a better strategy for the SU in re-rendezvous stage is to try to access the previously used channel instead of random access. Suppose there are $l_k$ re-rendezvous SUs at the beginning of slot $k$. One of the re-rendezvous SUs, SU $m_r$, accesses its previously utilized channel $n_r$ with probability $p_{m_r, n_r}=1$. The access distribution vector for SU $m_r$ becomes
\begin{align}
\mathbf{p}_{k,m}= \left[0, 0, \dots, 1,\dots,0 \right]^T,
\end{align}
where 1 is the $n_r$th element of the vector. There are $N_k-l_k$ initial-state SUs, and the SUs transmit on each available channel with probability vector
\begin{align}
\mathbf{p}_k=\left[\frac{1}{N_k}, \frac{1}{N_k}, \dots,   \frac{1}{N_k} \right].
\end{align}
The probability matrix $\mathbf{P}_k$ is updated to be $\mathbf{P}_k^{\prime}$ accordingly.

The expected number of successful transmissions without collision in channel $n_r$ is
\begin{align}
{\mathbb{E}}\left[ X_{n_r} \right] = {\left(1-\frac{1}{N_k}\right)}^{N_k-l_k}.
\end{align}
For the $n$th channel without re-rendezvous SUs, the expected number of successful transmissions without collision is
\begin{align}
{\mathbb{E}}\left[ X_{n} \right] = \frac{M_k-l_r}{N_k} {\left(1-\frac{1}{N_k}\right)}^{N_k-l_k-1}.
\end{align}
The total expected number of successful transmissions without collision during slot $k$ is
\begin{align}
Y^{\prime}\left( \mathbf{P}_k^{\prime} \right)=
&l_k {\left(1-\frac{1}{N_k}\right)}^{N_k-l_k} \nonumber \\
&+ \frac{\left(N_k-l_k\right)  \left(M_k-l_r\right)}{N_k} {\left(1-\frac{1}{N_k}\right)}^{N_k-l_k-1},
\label{eq:total-throughput-in-a-slot-simplified-re-rendezvous}
\end{align}
where $2\le M_k \le N_k$.

\begin{mytheorem}
\label{theorem:max-throughput-re-rendezvous}
$Y^{\prime}\left( \mathbf{P}_k^{\prime} \right)$ in (\ref{eq:total-throughput-in-a-slot-simplified-re-rendezvous}) and $Y\left( \mathbf{p}_k \right)$ in (\ref{eq:total-throughput-in-a-slot-simplified-max}) meet
\begin{align}
Y^{\prime}\left( \mathbf{P}_k^{\prime} \right) \ge Y\left( \mathbf{p}_k \right),\quad l_k \ge 0,
\end{align}
where $2\le M_k \le N_k$, and equality holds only if $l_k=0,1$.
\end{mytheorem}
The proof of Theorem~\ref{theorem:max-throughput-re-rendezvous} is in Appendix~\ref{appendix:proof-theorem-max-throughput-re-rendezvous}. Theorem~\ref{theorem:max-throughput-re-rendezvous} shows that a better strategy for a re-rendezvous SU is to try to access its previously utilized channel if $M_k \le N_k$.

\section{MAC Algorithms Design}
\label{sec:MAC-Algorithms-Design}
Based on the three types of SU information and the slot-level analysis, we explore three MAC algorithms based upon multichannel CSMA: Multichannel CSMA-F uses full SU information, Multichannel CSMA-P using partial SU information, and Multichannel CSMA with no SU information.

\subsection{Multichannel CSMA-F}
Algorithm~\ref{algorithm:MultichannelCSMA-F} specifies  Multichannel CSMA-F for an SU. During the initial state, an SU tries to access the spectrum based on (\ref{eq:access-probability-initial-stage}) to reduce the collision rate. During re-rendezvous, an SU tries to access its previously utilized channel if it is available and $M_k \le N_k$ as shown in Theorem~\ref{theorem:max-throughput-re-rendezvous}.

\begin{algorithm}[tb]
\caption{Multichannel CSMA-F}
\label{algorithm:MultichannelCSMA-F}
\begin{algorithmic}[1]
\Procedure{}{}
\State check data arrival
\If {no data arrived}
    \State Go to Step 2
\EndIf
\State do wideband spectrum sensing for $T_s$ on Tx side
\If {there is no available channel}
    \State wait a random backoff time $T_b$
    \State go to Step 6.
\EndIf
\If {$M_k \le N_k$, there is a previous transmission channel, and the previous channel is available}
    \State stay on the previous channel
\Else
    \State randomly choose one of the available channels and access it with probability $\min \{\frac{1}{M_k}, \frac{1}{N_k}\}$
\EndIf
\State transmit data for $T_d$
\State go to Step 2
\EndProcedure
\end{algorithmic}
\end{algorithm}

\begin{algorithm}[thb]
\caption{Multichannel CSMA-P}
\label{algorithm:MultichannelCSMA-F}
\begin{algorithmic}[1]
\Procedure{}{}
\State check data arrival
\If {no data arrived}
    \State Go to Step 2
\EndIf
\State do wideband spectrum sensing for $T_s$ on Tx side
\If {there is no available channel}
    \State wait a random backoff time $T_b$
    \State go to Step 6.
\EndIf
\If {$M_k < N_k$, there is a previous transmission channel, and the previous channel is available}
    \State stay on the previous channel
\Else
    \State randomly choose one of the available channels and access it with probability $\frac{1}{N_k}$
\EndIf
\State transmit data for $T_d$
\State go to Step 2
\EndProcedure
\end{algorithmic}
\end{algorithm}

\begin{algorithm}[tbh]
\caption{Multichannel CSMA}
\label{algorithm:MultichannelCSMA}
\begin{algorithmic}[1]
\Procedure{}{}
\State check data arrival
\If {no data arrived}
    \State Go to Step 2
\EndIf
\State do wideband spectrum sensing for $T_s$ on Tx side
\If {there is no available channel}
    \State wait a random backoff time $T_b$
    \State go to Step 6.
\EndIf
\State randomly choose one of the available channels
\State transmit data for $T_d$
\State go to Step 2
\EndProcedure
\end{algorithmic}
\end{algorithm}

%
%

Collisions occurs if two or more SUs sense during the same slot and select the same channel for transmission. A spectrum opportunity could be wasted if an SU completes its packet transmission in a slot while other SUs are conducting spectrum sensing. Full SU information decreases the collision probability in both initial state and re-rendezvous state.

\subsection{Multichannel CSMA-P}

Algorithm~\ref{algorithm:MultichannelCSMA-F} specifies  Multichannel CSMA-P for an SU. During re-rendezvous, an SU tries to access its previously utilized channel if it is available and $M_k < N_k$ as shown in Theorem~\ref{theorem:max-throughput-re-rendezvous}. Partial SU information decreases the collision probability in the re-rendezvous state.

\subsection{Multichannel CSMA}

Algorithm~\ref{algorithm:MultichannelCSMA} specifies  Multichannel CSMA for an SU. No SU information can be applied to improve the performance.

\subsection{Upper Bound on the Spectrum Sharing Efficiency}
%

The upper bound on the spectrum sharing efficiency can be obtained when there is no collision and no backoff and each SU has full data rate. Then
\begin{align}
E &\le \left\{ {\begin{aligned}
                {\frac{{\mathbb{E}\left[ {{T_d}} \right]}}{{{T_s} + \mathbb{E}\left[ {{T_d}} \right]}},\quad M \le N}  \\
                {\frac{{N\mathbb{E}\left[ {{T_d}} \right]}}{{M\left( {{T_s} + \mathbb{E}\left[ {{T_d}} \right]} \right)}},M > N}  \\
                \end{aligned}}
 \right. & \nonumber \\
&={\min \left( {1,\frac{N}{M}} \right)}{\frac{{\mathbb{E}\left[ {{T_d}} \right]}}{{{T_s} + \mathbb{E}\left[ {{T_d}} \right]}}}, \nonumber \\
&=E_{upper},
\end{align}
where $T_s$ is time for spectrum sensing, $ T_d $ is transmission time of a packet, and $\mathbb{E}$ denotes expectation.

This analysis assumes that $T_s$ is sufficiently large to make the effects of false alarms and missed detections from spectrum sensing negligible.
Decreasing $T_s$ increases the upper bound on the spectrum sharing efficiency; however, it would also increase the probabilities of false alarm and miss detection. False alarms would decrease the spectrum sharing efficiency because an SU may be waiting for spectrum while there is available spectrum. Miss detections would increase the probability of collision, decreasing the spectrum sharing efficiency. On the other hand, increasing ${\mathbb{E}\left[ {{T_d}} \right]}$ would also increase the spectrum sharing efficiency, but the delay for accessing channels would also increase for $M>N$ in this case. More detailed performance analysis for the slotted case is provided in \cite{mingmingcaiJNLAllerton2015}.

%

Let $E_f$, $E_p$, and $E_n$ be the the spectrum sharing efficiencies for Multichannel CSMA-F, Multichannel CSMA-P, and Multichannel CSMA, respectively. We expect them to have the following relation ordering
\begin{align}
E_n \le E_p \le E_f \le E_{upper}.
\end{align}
The remainder of this suggestion illustrates these relationships more specifically through simulation.

\subsection{Simulation of MAC algorithms}
\label{sec:Simulation-of-MAC-algorithms}

We simulate and compare the three MAC algorithms. The simulation follows the system architecture. No PU is considered, and we assume that the total number of channels available for SUs to access remains the same during simulation.
Backoffs follows a geometric distribution with mean $10$ slots. The packet size is measured by the transmission time of the packet, and it has uniform distribution $U \left( T_{d, min}, T_{d, max} \right)$ in slots. The average packet size
\begin{align}
{\mathbb{E}\left[ {{T_d}} \right]} =\frac{T_{d, min}+T_{d, max}}{2}.
\end{align}
If $T_{d, min}= T_{d, max}$, the packet has fixed size. Packet arrivals follow a Poisson process with mean arrival interval $\lambda$ slots. The smaller $\lambda$ is, the smaller the expected interval between two packets will be, and the more packets will arrive in a given period. For the purposes of simulation, the three MAC algorithms have exactly the same packet sizes and arrival times.

\begin{figure}
\centering
\includegraphics[width=90 mm]{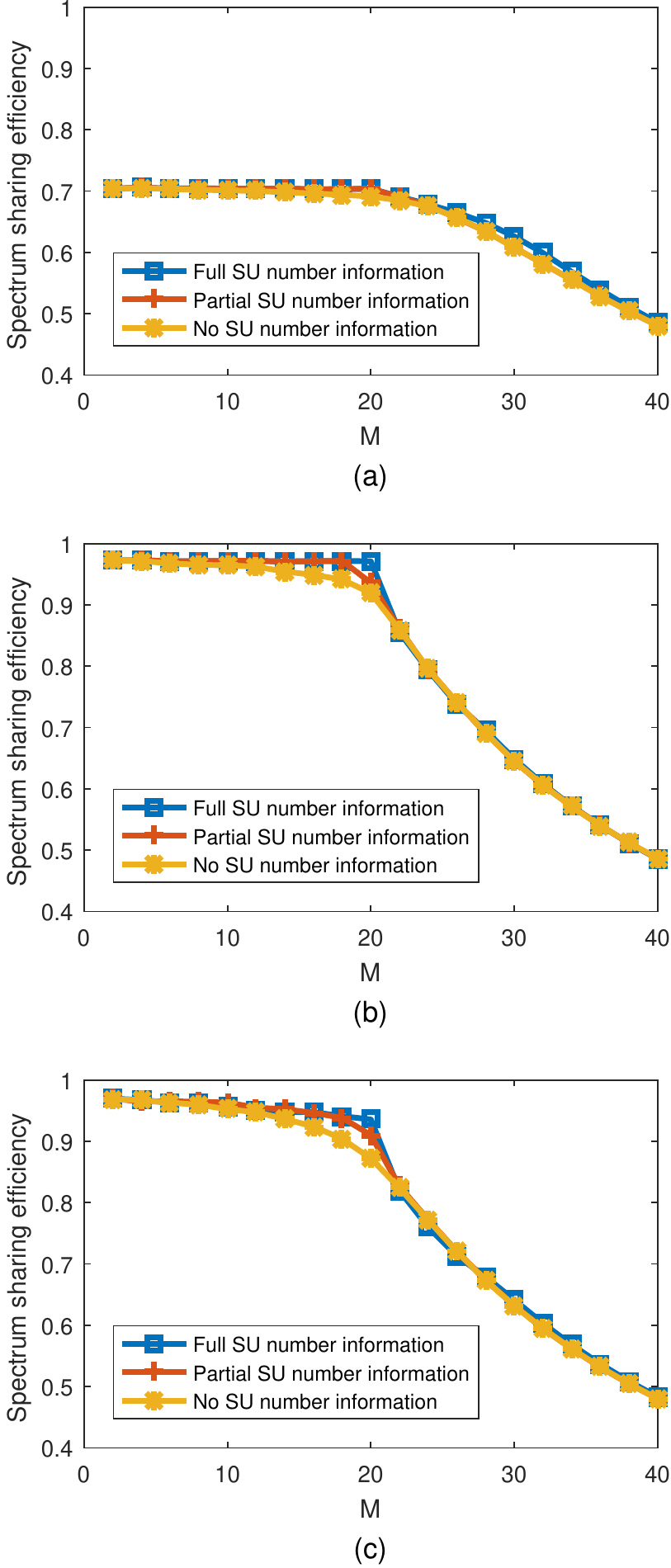}
\caption{Simulation of spectrum sharing efficiency as a function of the number of users $M$ for fixed number of channels $N=20$ and fixed packet size of 50 slots. Packet arrivals follow a Poisson process with mean arrival interval $\lambda$ slots. (a) $\lambda=70$. (b) $\lambda=50$. (c) $\lambda=20$.}
\label{fig:SSE_packet_size_50_50}
\end{figure}

Figure~\ref{fig:SSE_packet_size_50_50} displays the simulation results for packets with fixed packet size of $50$ slots, for $\lambda=70$, $\lambda=50$, and $\lambda=20$, respectively. The simulation results demonstrate that $E_n \le E_p \le E_f $. However, the spectrum sharing efficiency is not significantly improved by the number of SUs, and the largest improvement apears to occur if $M$ is around $N$. Comparing Figure~\ref{fig:SSE_packet_size_50_50} (a) to \ref{fig:SSE_packet_size_50_50} (b) , the packet arrival interval decreases, ${\mathbb{E}\left[ {{T_d}} \right]}<\lambda$  changes to ${\mathbb{E}\left[ {{T_d}} \right]}=\lambda$, and the spectrum sharing efficiencies for all MAC algorithms increase because there are fewer idle slots without packets for transmission.  Comparing Figure~\ref{fig:SSE_packet_size_50_50} (b) to \ref{fig:SSE_packet_size_50_50} (c), the packet arrival interval decreases even further, ${\mathbb{E}\left[ {{T_d}} \right]}=\lambda$  changes to ${\mathbb{E}\left[ {{T_d}} \right]}>\lambda$, and the spectrum sharing efficiencies for all MAC algorithms slightly decrease because of higher collision rate with overloaded packets.

\begin{figure}
\centering
\includegraphics[width=90 mm]{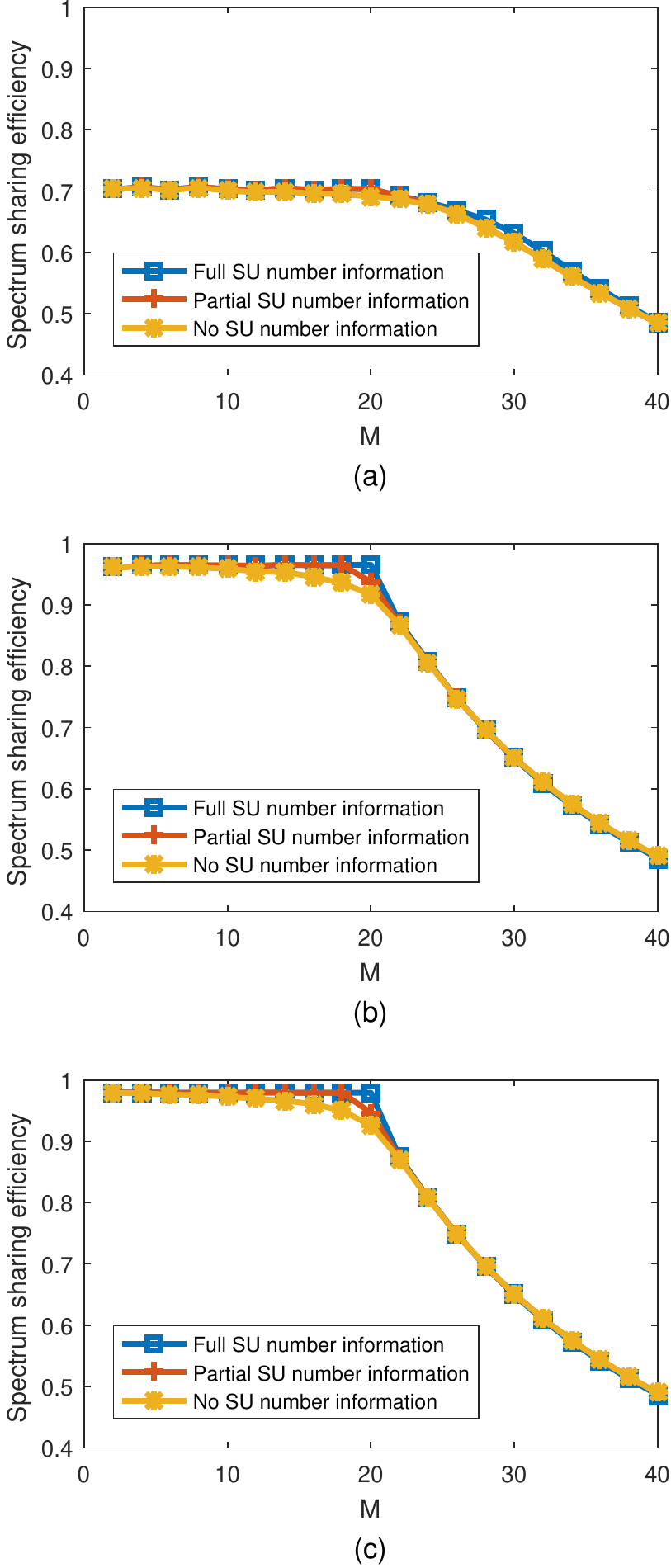}
\caption{Simulation of spectrum sharing efficiency as a function of the number of users$M$ for fixed number of channels $N=20$. Packet sizes have a uniform distribution $U \left( 30, 70 \right)$ slots. Packet arrivals follow a Poisson process with mean arrival interval $\lambda$ slots. (a) $\lambda=70$. (b) $\lambda=50$. (c) $\lambda=20$.}
\label{fig:SSE_packet_size_30_70}
\end{figure}

Figure~\ref{fig:SSE_packet_size_30_70} illustrates the simulation results for packet sizes with distribution $U \left( 30, 70 \right)$ for $\lambda=70$, $\lambda=50$, and $\lambda=20$, respectively. We can observe similar trends as those of Figure~\ref{fig:SSE_packet_size_50_50}.

Figure~\ref{fig:SSE_packet_size_50_50} (c) and \ref{fig:SSE_packet_size_30_70} (c) have the same average packet size and average packet arrival interval. However, the spectrum sharing efficiencies in Figure~\ref{fig:SSE_packet_size_30_70} (c)  are higher than those of Figure~\ref{fig:SSE_packet_size_50_50} (c). The reason is that randomized packet sizes reduce the probability of collision to some extent.

\subsection{Discussion}
Our analysis suggests that SU information only slightly improves the spectrum sharing efficiency for multichannel CSMA-based MAC protocols, and the largest improvement occurs for $M$ around $N$. However, to provide the SU information, the system requires either spectrum sensing during transmission or collection of SU information by the database. On one hand, sensing during transmission requires more sophisticated and expensive transceivers; on the other hand, acquiring full SU information requires a third-level database, which we expect to be much more costly than a second-level database.  From these perspectives, Multichannel CSMA without SU information has advantages over both Multichannel CSMA-F and Multichannel CSMA-P.  We therefore focus on Multichannel CSMA for implementation in an advanced SDR platform, as addressed in the next section.

\section{SDR Prototyping}
\label{sec:SDR-Prototyping-of-Multichannel-CSMA}
This section overviews our SDR prototype of wideband distributed spectrum sharing as it would be implemented in a SU device. The prototype includes a spectrum analyzer, a faster-rendezvous physical layer, and Multichannel CSMA.

We leverage an SDR platform acquired from National Instruments that includes the USRP RIO 2953R hardware and LabVIEW CSDS and its associated LTE Application Framework.  This platform was chosen for prototyping because the existing physical layer provided a number of baseline features that we can repurpose for MIMA, specifically, orthogonal frequency division multiplexing (OFDM) with the ability to turn subcarriers on and off, a variety of built-in modulation and coding schemes (MCS), carrier and timing synchronization, and so forth.

In the prototype, a 20~MHz bandwidth is channelized into 10 MIMA channels.  The selection of 20~MHz and 10 channels is limited by the currently available hardware, but could be expanded with upgrades in front-end bandwidth and signal processing capabilities. Those familiar with LTE know that it uses OFDM subcarrier spacing of 15~kHz and groups twelve OFDM subcarriers together into a so-called PRB. We group ten PRBs together to create a PRB Group (PRBG), one for each MIMA channel.
For sensing and Multichannel CSMA, we use a slot that is 140 OFDM symbols / one LTE frame in duration.  Although the implementation is slotted, it can approximate a continuous system if the average transmission of a data packet is much larger than the slot length.

%
%

\subsection{Spectrum Analyzer}

Among the spectrum sensing methods, energy detection requires limited prior knowledge about the signals of interest. An FFT-based spectrum analyzer has been developed for wideband spectrum sensing with the USRP RIO and NI CSDS. The spectrum analyzer is designed to detect the availability of each of the 10 MIMA channels.

\begin{figure*}
\centering
\includegraphics[width=140mm]{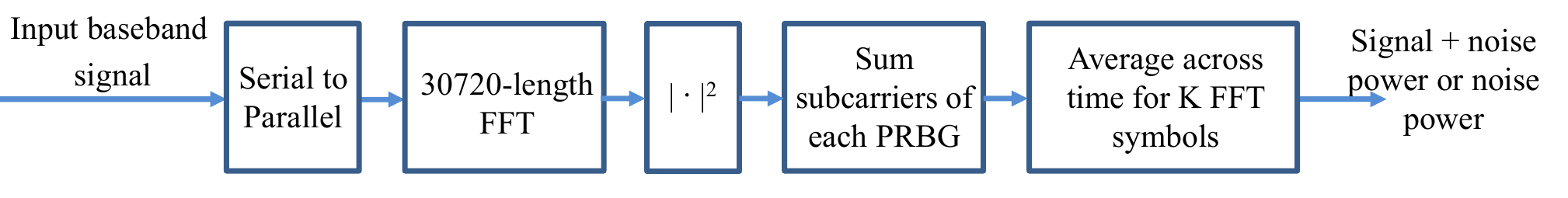}
\caption{Signal preprocessing for the spectrum analyzer.}
\label{spectrum-analyzer}
\bigskip
\centering
\includegraphics[width=180mm]{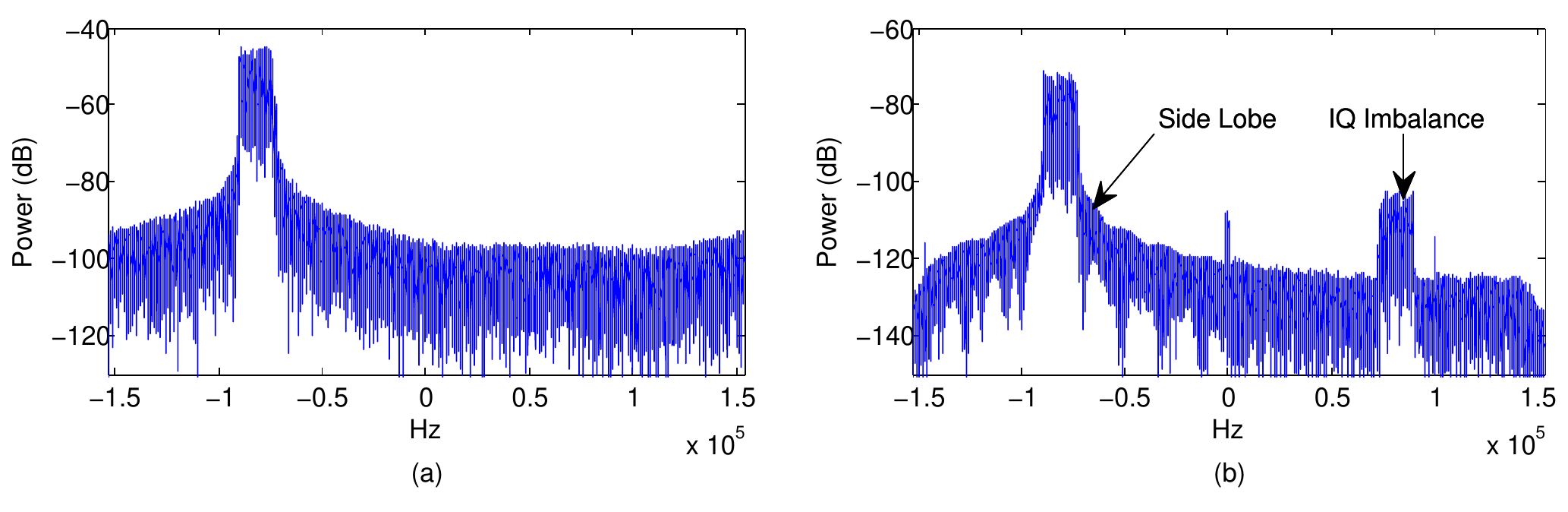}
\caption{An example of side lobe and IQ imbalance observed by the demodulator of the receiver. One PRBG is used for data transmission. (a) transmitted spectrum (b) received spectrum.}
\label{IQ_imbalance_TxRx}
\end{figure*}

The signal processing is illustrated in Figure~\ref{spectrum-analyzer}.  This signal processing is conducted in a general-purpose processor.  Four steps are taken to process the received data. First, a 30720-length FFT is calculated to convert the time domain signal to frequency domain.  The number of samples of an LTE subframe is 30720, including the cyclic prefix. The FFT size of 30720 is selected to save computational resources, instead of computing a number of smaller FFTs and then averaging over time. Let $x_{i,k},\ i=1,2,\dots, 30720$ be the received frequency samples after the FFT, and $w_{i,k},\ i=1,2,\dots, 30720$ be the noise power, where $k=1,2,\dots,K$ denotes the index of FFT symbol. Second, we calculate the power of each frequency sample ${\left| x_{i,k} \right|}^2$. Third, for each PRBG, we sum all the frequency samples within that PRBG. Fourth, we average the power of each PRBG across time for $K$ FFTs. In our implementation, $K=10$ is used to average the signal with length of one LTE frame, i.e., 10 subframes.  We note that no timing synchronization with the LTE frame is required for the spectrum analyzer.

The output of the signal processing block for PRBG $n$ would be
\begin{align}
{y_n} = \frac{1}{K}\sum\limits_{k = 1}^K {{\sum _{i \in {{\cal D}_j}}}{{\left| {{x_{i,k}} + {w_{i,k}}} \right|}^2}} ,n = 1,2, \ldots ,N,
\end{align}
where $N$ is the number of PRBGs and $\mathcal{D}_n$ is the set of FFT samples belonging to PRBG $n$. Each PRBG can be detected separately. For PRBG $n$, the hypotheses would be
\begin{align}
&\mathcal{H}_{0,n} \colon {y_n} = \frac{1}{K}\sum\limits_{k = 1}^K {{\sum _{i \in {{\cal D}_j}}}{{\left| {w_{i,k}} \right|}^2}}, &\\
&\mathcal{H}_{1,n} \colon {y_n} = \frac{1}{K}\sum\limits_{k = 1}^K {{\sum _{i \in {{\cal D}_j}}}{{\left| {{x_{i,k}} + {w_{i,k}}} \right|}^2}}.
\end{align}
Denote the probability of detection, the probability of false alarm, and the probability of miss detection for PRBG $n$ as ${P_{D,n}}$, ${P_{F,n}}$ and ${P_{M,n}}$, respectively. Then
\begin{align}
&{P_{D,n}} = \Pr \left( {{y_n} > \gamma |{\mathcal{H}_{1,n}}} \right), &\\
&{P_{F,n}} = \Pr \left( {{y_n} > \gamma |{\mathcal{H}_{0,n}}} \right), & \\
&{P_{M,n}} = \Pr \left( {{y_n} \le \gamma |{\mathcal{H}_{1,n}}} \right)=1-{P_{D,n}} ,
\end{align}
where $\gamma$ is the threshold for detection. For a fixed SNR, as $\gamma$ increases,  ${P_{M,n}}$ increases, and ${P_{D,n}}$ and ${P_{F,n}}$ decreases. In a distributed spectrum sharing system, miss detections usually have higher cost than false alarms. Therefore, $\gamma$ should be chosen to be relatively small. As SNR increases, ${P_{F,n}}$ and ${P_{M,n}}$ decrease and ${P_{D,n}}$ increases with a properly chosen $\gamma$.

In practice, false alarms are also caused by signal side lobs as well as images that result from in-phase (I) and quadrature (Q) imbalance.  Usually, the first side lobe adjacent to the signal in frequency domain has the highest energy. For intermediate frequency reception, IQ imbalance causes the well-known image problem as shown in Figure~\ref{IQ_imbalance_TxRx} \cite{mailand2006iq}. The imaged signal of IQ imbalance varies with different USRPs.

In this paper, we focus on demonstration of the whole architecture. We assume SNR is high, such as more than 20 dB, to guarantee data success rate. In this case, there is more room to choose the threshold so that ${P_{F,n}}$ and ${P_{M,n}}$ are small.

\subsection{Continuous-Time Multichannel CSMA}
\begin{figure}
\centering
\includegraphics[width=90mm]{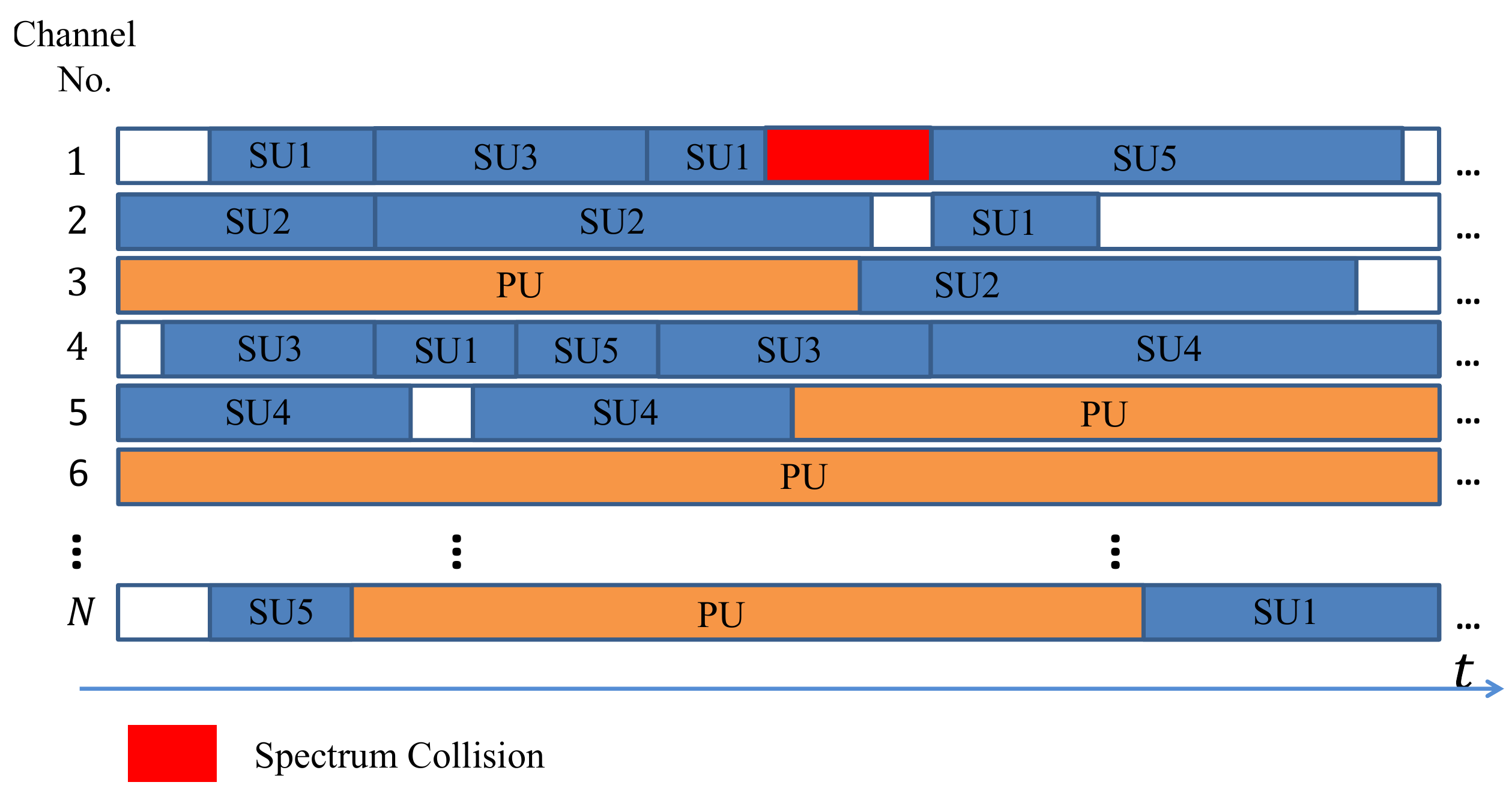}
\caption{An example of Multichannel CSMA. A spectrum collision happens because SU5 has a missed detection error.}
\label{MultiChannel_CSMA_example}
\end{figure}

The continuous-time Multichannel CSMA Algorithm for any SU to access available channels is similar to Algorithm~\ref{algorithm:MultichannelCSMA}. Figure~\ref{MultiChannel_CSMA_example} illustrates Multichannel CSMA in action.
Collisions occur if two or more SUs transmit on the same channel at the same time. If the database is assumed to be perfect, the overall spectrum sharing efficiency depends on three components, the physical layer, spectrum sensing, and the MAC algorithm. Imperfections in any of the three components could degrade system performance.

For example, a defective physical layer causes lower data rate than expected. Signal distortion in hardware caused by IQ imbalance, nonlinearities, and quantization noise could degrade the physical layer performance. If spectrum sensing has a high probability of false alarm or miss detection, the system would waste spectrum opportunities or have a large number of collisions. An inefficient channel assignment protocol reduces spectrum opportunities available to SUs. Even in the case of low probability of false alarm and miss detection for spectrum sensing, collisions can also occur if two or more users do spectrum sensing simultaneously and then decide to transmit on the same channel.

\subsection{Experimental Results}
Two experiments have been designed and conducted to validate the spectrum analyzer and the whole system, including the fast-rendezvous physical layer, spectrum analyzer and channel assignment protocol. Test equipments are NI USRP RIO 2953R and NI LabVIEW CSDS. The center frequency is 2.4 GHz for all experiments.

\subsubsection{Experimental Results of Spectrum Analyzer}
The selection of thresholds should minimize the probability of false alarm and miss detection. The test is over the air through antennas. The maximum SNR that can be obtained in our setup is about 45dB. Tx and Rx ports in the same USRP are run as an SU pair for the test. The signal power is maximized in this case. PRBG 1 is chosen as an example. The false alarm ratio is defined as the ratio of the number of false alarms to the total number of tests. Similarly, the miss detection ratio is defined as the ratio of the number of miss detections to the total number of tests. As the test time $t \to \infty$, the false alarm rate $\to P_F$, and the miss detection rate $\to P_M$.
\begin{figure}[t]
\centering
\includegraphics[width=85mm]{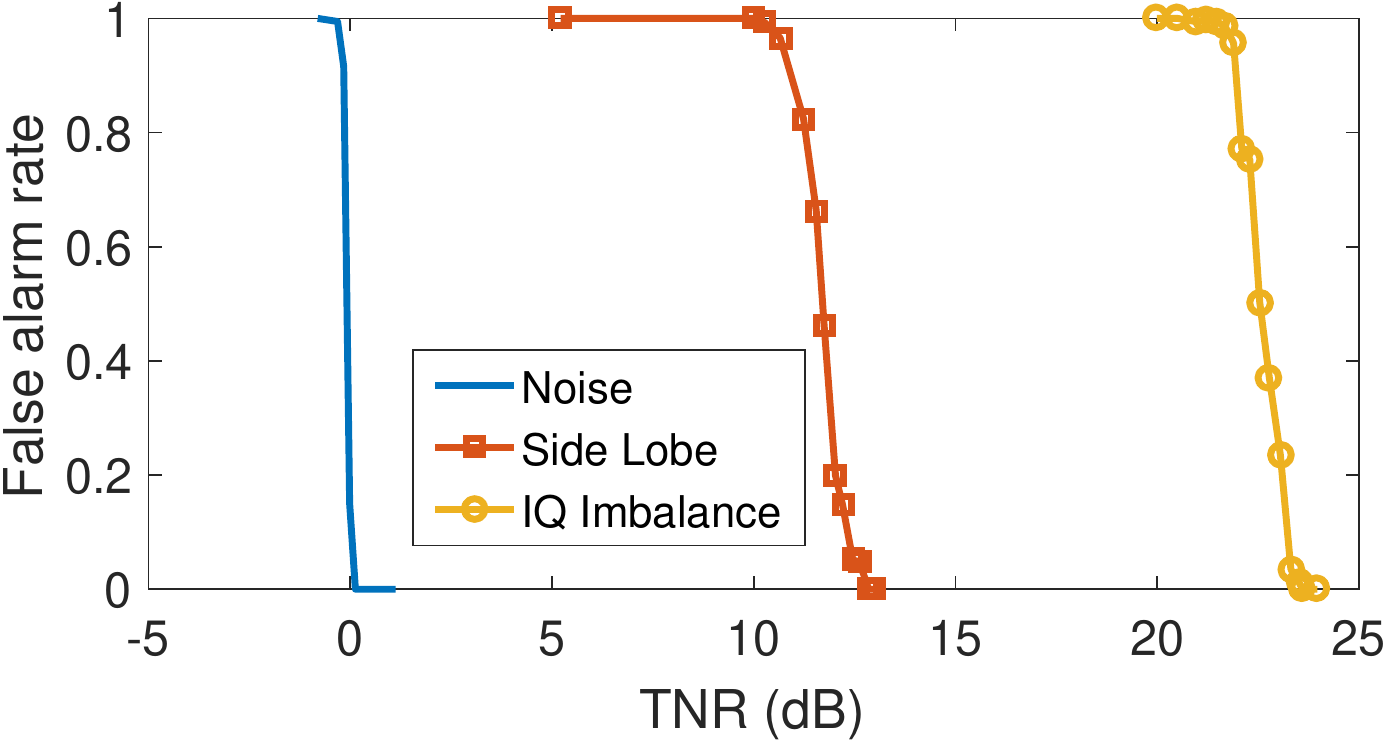}
\caption{The false alarm rate for noise, largest side lobe signal, and IQ signal imbalance as a function of threshold-to-noise ratio (TNR). The test is conducted on the USRP device with the worst IQ imbalance performance. $K=10$.}
\label{SpectrumSensing_PF_noise_sidelobe_IQ}
\end{figure}

\begin{figure}[t]
\centering
\includegraphics[width=85mm]{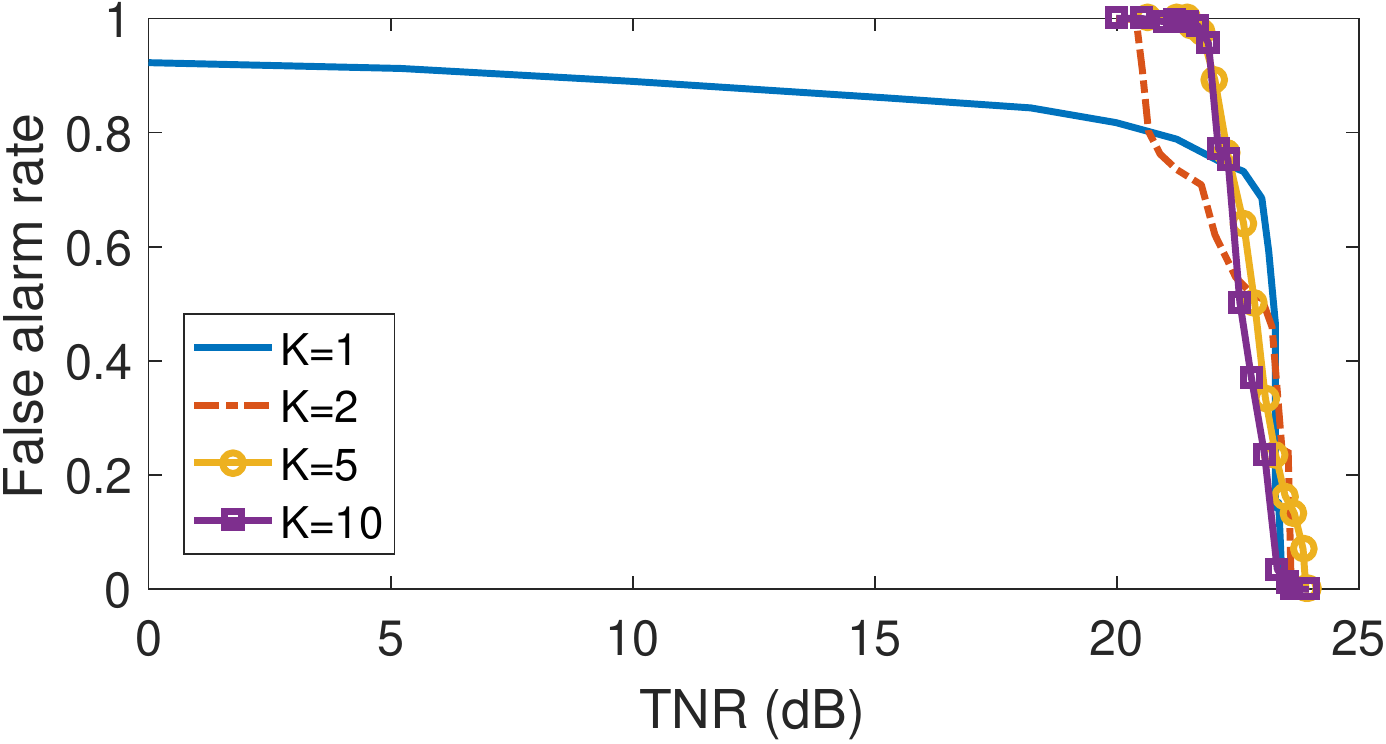}
\caption{The false alarm rate of IQ imbalance signal as a function of TNR. The test is conducted on the USRP device with the worst IQ imbalance performance. $K=1,2,5,10$.}
\label{SpectrumSensing_PF_IQ_Vary_L}
\end{figure}

\begin{figure}[th]
\centering
\includegraphics[width=85mm]{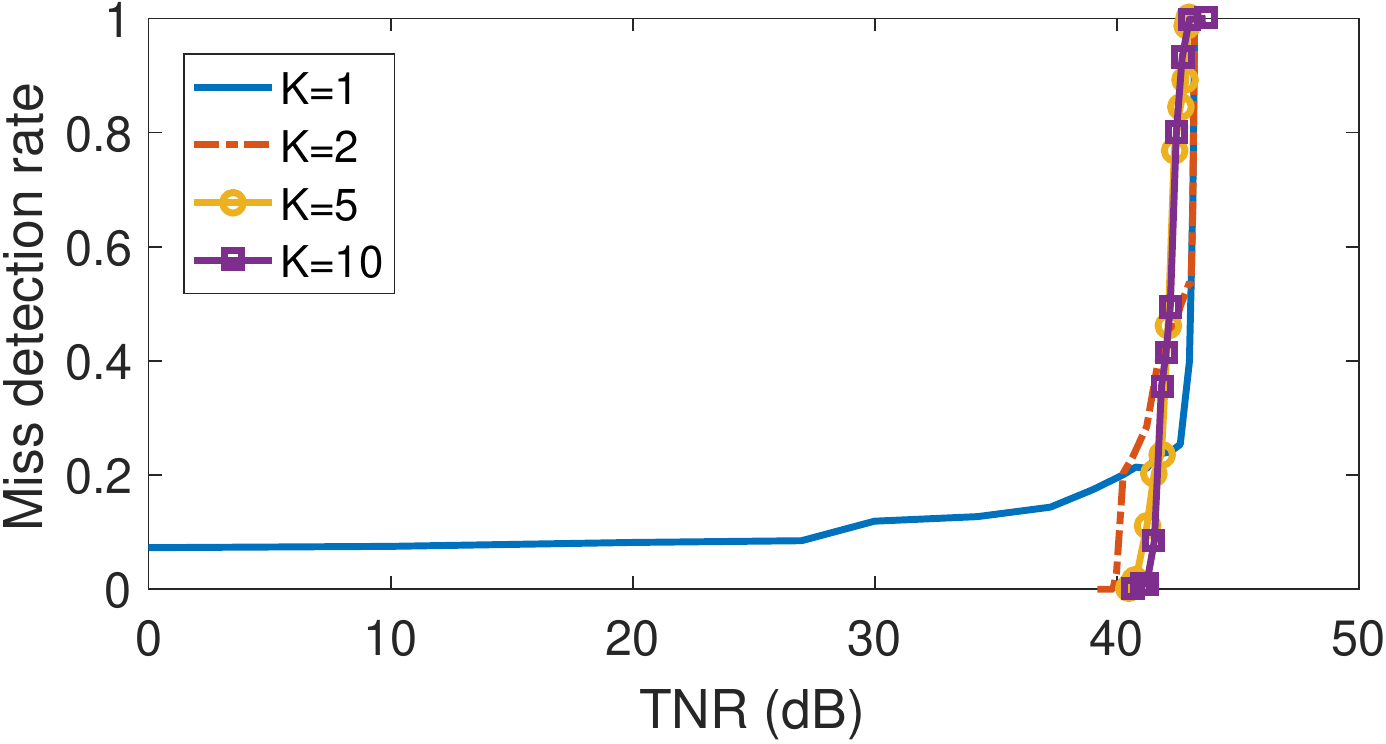}
\caption{The miss detection rate as a function of TNR. SNR$=42$ dB. $K=1,2,5,10$.}
\label{SpectrumSensing_PM}
\end{figure}

In Figure~\ref{SpectrumSensing_PF_noise_sidelobe_IQ}, the false alarm rate is tested for noise, signal from the adjacent side lobe, and IQ imbalance image as a function of threshold-to-noise ratio (TNR). The false alarm rate for noise is obtained when Tx is turned off. The false alarm rate for the side lobe is tested on PRBG 2 while PRBG 1 is turned on for data transmission. The false alarm rate for the IQ imbalance image is tested on PRBG 10 while the data signal is transmitted on PRBG 1. PRBG 10 and PRBG 1 are symmetric with respect to DC. We choose the USRP with the worst IQ imbalance performance for the test to highlight the effects of IQ imbalance on spectrum sensing. The power of the IQ imbalance image is about 22.5 dB higher than the noise, while the side lobe power is about 12 dB higher than the noise. From the test results, the threshold should be several dB higher than the IQ signal imbalance image to make sure the false alarm rate is close to 0.

In Figure~\ref{SpectrumSensing_PF_IQ_Vary_L}, we measure the false alarm rate of the IQ imbalance image as a function of TNR for different $K$. Again, as shown in Figure~\ref{spectrum-analyzer}, $K$ is the number of FFT symbols to calculate the moving average across time for spectrum sensing. The test results show that the false alarm rate decreases quickly from 1 to 0 if the threshold is about 1.5 dB higher than the IQ imbalance image. If $K=1$, the signal to the spectrum analyzer occasionally comes from UL subframe, where there is no signal currently. This explains why false alarm curve for $K=1$ is less than 1 even for a small threshold.

In Figure~\ref{SpectrumSensing_PM}, the signal miss detection rate is tested as a function of TNR for different $K$. If $K=1$, the miss detection is always larger than 0, because of the empty UL subframe. If $K \ge 2$, the miss detection is zero when the threshold is about 2 dB smaller than the signal power.

In summary, $K \ge 2$ is required. To avoid false alarm, the threshold should be at least 1 dB higher than the maximum of noise power, side lobe power and power of the IQ imbalance image. The threshold should be at least 2 dB smaller than the signal power to avoid miss detection. IQ imbalance comes from the hardware, and it varies for different USRPs. If the channel with IQ imbalance signal is considered as not available, the spectrum opportunity is wasted. On the other hand, if the channel with the IQ imbalance image is considered as available, the IQ imbalance image could interfere with the data signal and degrade the system performance, which will be apparent in the system tests.

$K=2$ corresponds to 2 ms of spectrum sensing. In practice, the radio needs almost 200 ms to transition between sensing and data transmission. Finally, we select $T_s=200$ ms and $K=10$. The sensing threshold is chosen to be 5 dB higher than the maximum side lobe power.

\subsubsection{Experimental Results for Spectrum Sharing Efficiency}
\begin{figure}[tb]
\centering
\includegraphics[width=85mm]{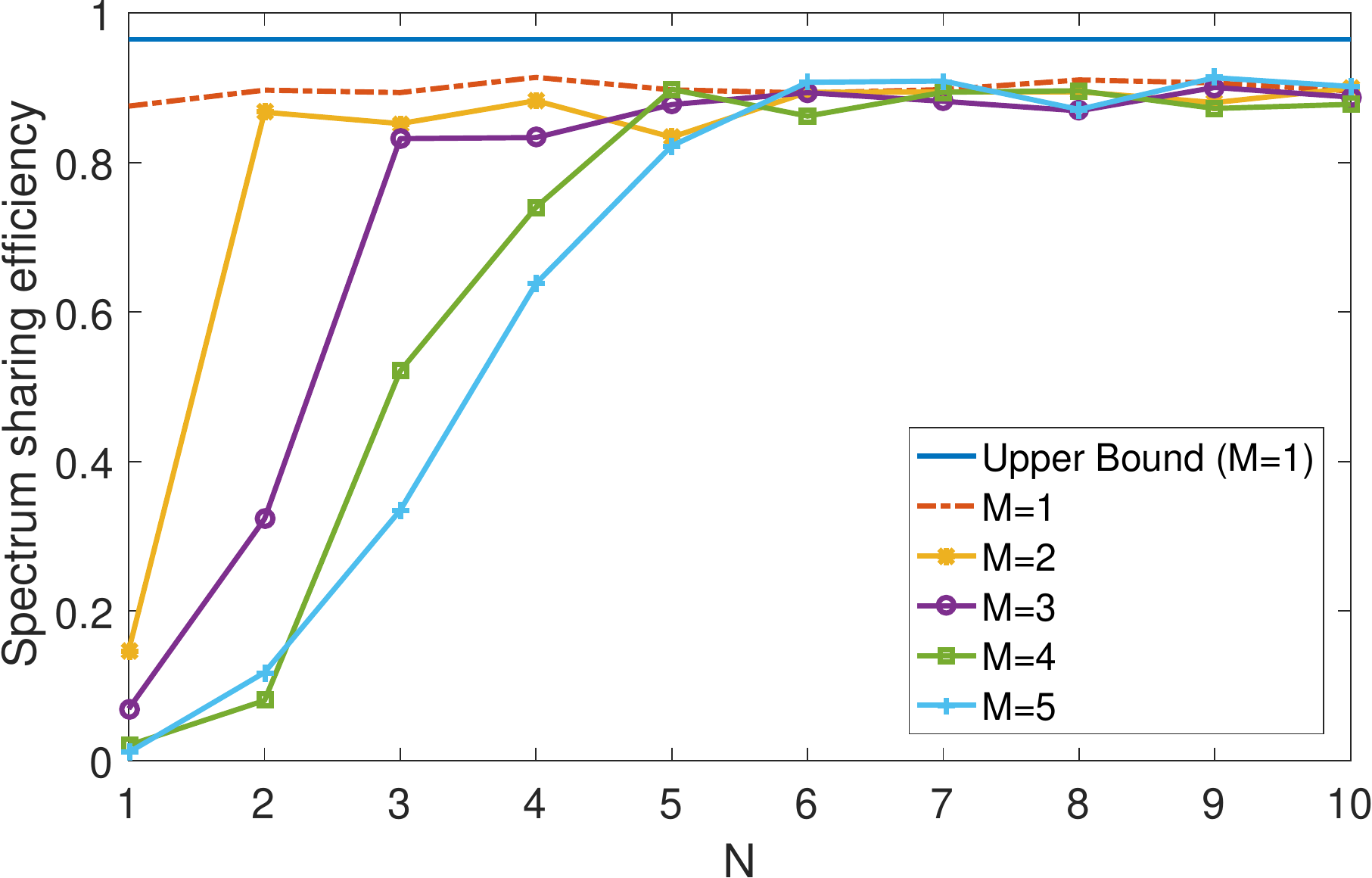}
\caption{Spectrum sharing efficiency as a function of the number of available channels $N$. Full data rate, LTE MCS$=13$ (16QAM and coding rate 0.48). $T_s=0.2$~s. $T_d$ and $T_b$ have uniform distributions with $T_d \sim U \left( 2s, 7s \right)$ and $T_b \sim U \left( 0, 0.2s \right)$. $M=1,2,3,4,5$. SNR=42 dB.}
\label{Efficiency_whole}
\end{figure}

\begin{figure}[tb]
\centering
\includegraphics[width=85mm]{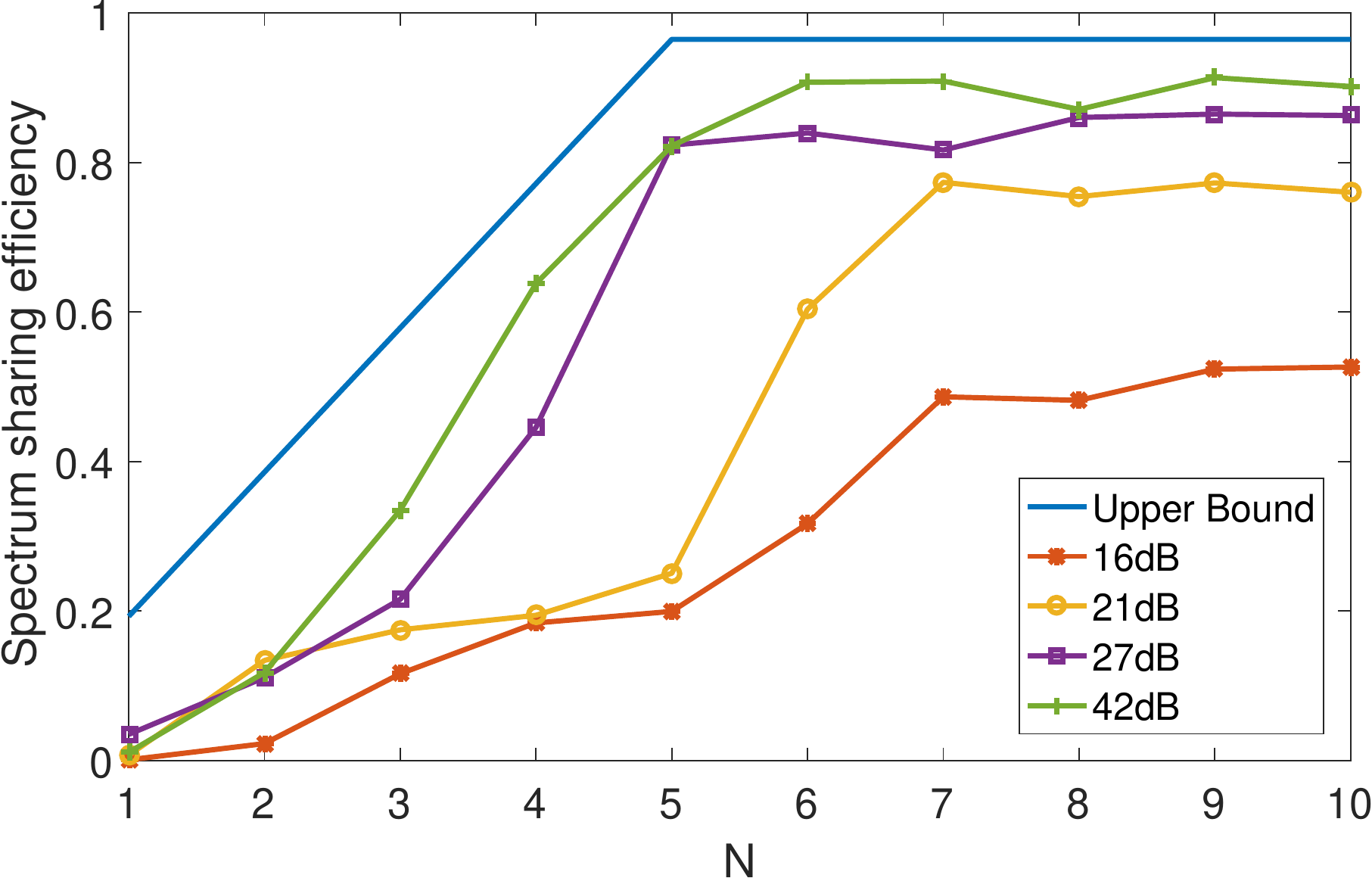}
\caption{Spectrum sharing efficiency as a function of the number of available channels $N$ for SNR=$16, 21, 27, 42$~dB. Full data rate, $M=5$ and MCS$=13$ (16QAM and coding rate 0.48). $T_s=0.2$~s. $T_d$ and $T_b$ have uniform distributions with $T_d \sim U \left( 2s, 7s \right)$ and $T_b \sim U \left( 0, 0.2s \right)$.}
\label{Efficiency_SNR}
\end{figure}

\begin{figure}[htb]
\centering
\includegraphics[width=85mm]{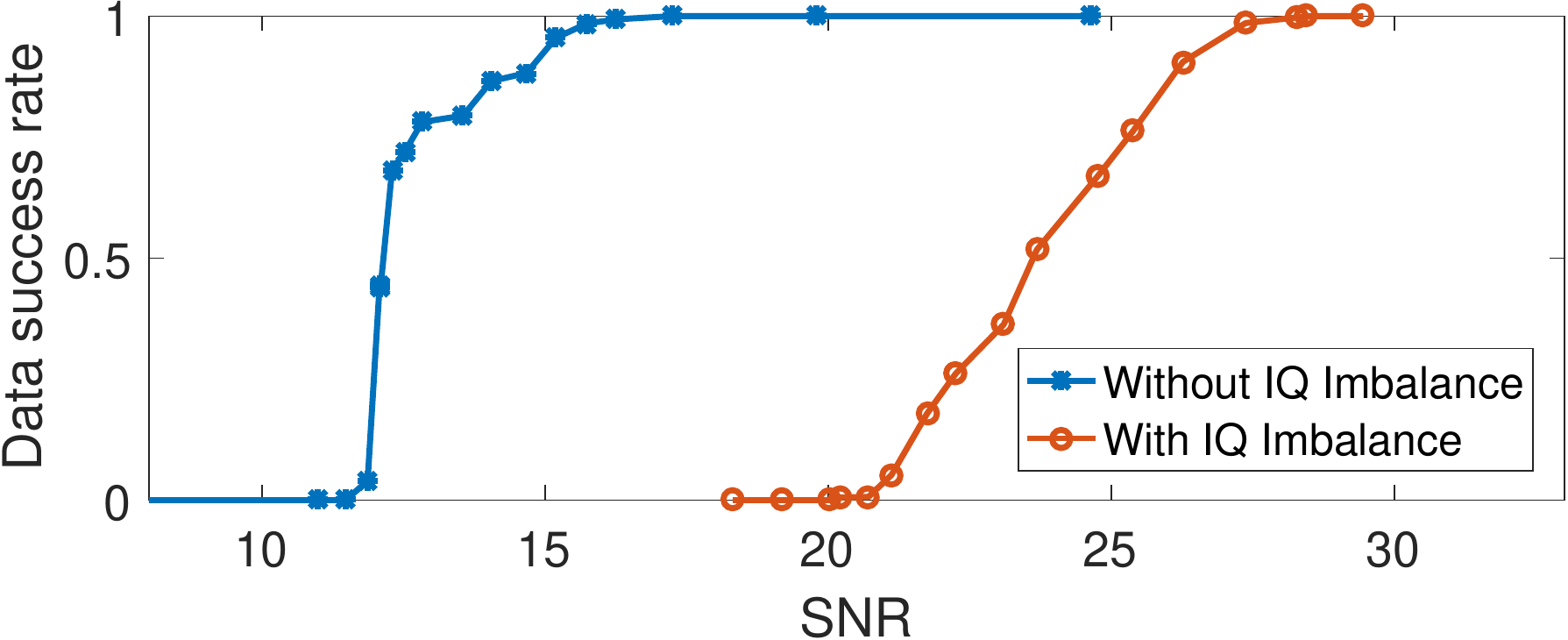}
\caption{Examples of data success rate with and without IQ imbalance causing interference. The IQ imbalance signal is 9.9~dB higher than noise, and the SNR of the signal that causes the IQ imbalance is about 42~dB. PRBG 1 is used for data transmission. The signal that causes IQ imbalance is transmitted on PRBG 10 so that the IQ imbalance image is on PRBG 1. MCS$=13$ (16QAM and coding rate 0.48).}
\label{IQ_imbalance_effect}
\end{figure}

For the purpose of demonstration, all the USRPs for transmitting are located closely to minimize the spectrum sensing error and to reduce the hidden node issue. The total number of available channels $N$ is manually selected so that a database is not necessary for these experiments. All SUs are assumed to have enough data to be active for data transmission all the time.

Figure~\ref{Efficiency_whole} illustrates the results for the spectrum sharing efficiency at high SNR (42 dB). There is gap between the upper bound and the actual measured spectrum sharing efficiency. This gap is caused by false alarms, miss detections, and collisions as well as transition between spectrum sensing and data transmission. The gap also exists if there is only one SU, showing that transition is the main overhead, which is something we are continuing to explore.
Overall, the system has better performance if $M \le N$. If $M>N$, collisions occur more frequently.

Figure~\ref{Efficiency_SNR} shows the spectrum sharing efficiency for several SNRs. As SNR increases, the spectrum sharing efficiency increases. If SNR is relatively small, such as 16~dB, the spectrum sharing efficiency is far from the upper bound. The physical layer with SNR 16~dB cannot obtain full data success rate without interference according to the tests in \cite{mingmingcaiJNLAsilomar2015}. The tests in \cite{mingmingcaiJNLAsilomar2015} also show that the physical layer can achieve full data rate if SNR$>17$~dB. However, there is a gap between the spectrum sharing efficiency of SNR$=21, 27$ dB and SNR$=$42 dB, which are all higher than the requirement, SNR$>17$ dB. False alarms and miss detections of the spectrum analyzer could be one reason. The interference of the IQ imbalance image may play a significant role.

Figure~\ref{IQ_imbalance_effect} illustrates an example of the increase of minimum SNR required for achieving full data rate because of the interference from IQ imbalance image. If the IQ imbalance signal is 9.9~dB, another 10~dB transmission power is required for full data success rate. If the spectrum analyzer treats the IQ imbalance as another SUs' signal, the spectrum opportunity is wasted. If the IQ imbalance is considered as noise by the spectrum analyzer, it  interferes with other SUs' data signals.

\subsection{Discussion of Software Design Issues}
The powerful FPGA in the USRP can significantly increase the capacity of signal processing compared to a host computer. In our MIMA physical layer, ten synchronization loops are implemented in the FPGA. The FPGA can still manage the large amount of signal processing. However, 96\% of the FPGA's resources has been used, preventing additional functionality and wider bandwidth.

In FPGA programming, the routing of logic gates is limited by the timing constraints of the logic gates. Timing constraint violations happen more frequently as the increase of the FPGA resource usage. The current synchronization algorithm and FPGA program require additional optimization to reduce the FPGA resource usage.

\section{Conclusion and Ongoing Work}
\label{sec:Conclusion-ongoing-work}
We developed a system architecture for a distributed spectrum sharing system. The requirements for the four components of the system, i.e., database, wideband spectrum sensing, fast-rendezvous physical layer, and MAC algorithm were discussed separately. We showed that the number of SUs could be used to improve the spectrum sharing efficiency. However, the improvement is smaller than the cost of acquiring the SU information. Therefore, Multichannel CSMA without the SU information is chosen for prototyping. The fast-rendezvous physical layer MIMA, a spectrum analyzer, and a Multichannel CSMA MAC algorithm have been developed based on the NI LabVIEW CSDS and USRP RIOs. We summarized elements of the design, and conducted experiments to validate the architecture of the wideband spectrum sharing radio. Tests showed that the radio is robust to out-of-band interference, and the radios can share the spectrum efficiently if the number of channels is larger than or equal to the number of users and when the SNR is high enough to overcome the interference from IQ imbalance.

More work is ongoing to improve the synchronization performance of the physical layer, to calibrate the IQ imbalance of specific USRPs, and to speed up transitions between spectrum sensing and data transmission. The whole system will also be optimized and tested in a larger network with dozens of radio nodes. We could also design and implement a database and integrate it into the system. Finally, the advantages of our spectrum sharing radios could combine with those of other radios to achieve better performance.

\section*{Acknowledgment}
This work has been partially supported by NSF Grant CCF11-17365. We thank J\"{o}rg Hofrichter, Amal Ekbal, Markus Unger and Ian Wong from National Instruments, who provided insight and expertise on the NI LTE Application Framework.

\renewcommand*\appendixpagename{\large APPENDICES}  
\begin{appendices}

\section{PROOF OF THEOREM 1}
\label{appendix:proof-theorem-max-throughput-initial}

The detailed system setup is illustrated in Section~\ref{sec:MAC-Layer-Analysis-with-Full-SU-Information}.

The partial derivative of $Y\left( \mathbf{p}_k \right)$ with respect to $p_n, n=1,2,\dots,N_k$ is
\begin{align}
\frac{\partial Y\left( \mathbf{p}_k \right) }{\partial p_n}= M_k \left(1-M_k p_n \right) \left( 1- p_n\right)^{M_k-2}.
\end{align}
Assume $M_k\ge2$. For any $n=1,2,\dots,N_k$
\begin{align}
\frac{\partial Y\left( \mathbf{p}_k \right) }{\partial p_n}=
\left\{ {\begin{array}{*{20}{c}}
   { > 0, \quad {p_n} < \frac{1}{{{M_k}}}}  \\
   { = 0, \quad {p_n} = \frac{1}{{{M_k}}}}  \\
   { < 0, \quad {p_n} > \frac{1}{{{M_k}}}}  \\
\end{array}} \right.
.
\end{align}

We divide further discussion into two scenarios, $M_k>N_k$ and $M_k \le N_k$. If $M_k>N_k$, $Y\left( \mathbf{p}_k \right)$ is maximized for ${p_n} = \frac{1}{M_k}$. Correspondingly,
\begin{align}
Y\left( \mathbf{p}_{k} \right) = {N_k} \left( 1- \frac{1}{M_k} \right)^{M_k-1}.
\end{align}
An SU does not transmit on any channel at slot $k$ with probability $1-\frac{N_k}{M_k}$.

If $M_k \le N_k$, all SUs need to transmit on a channel in order to maximize $Y\left( \mathbf{p}_{k} \right)$, and $\sum\limits_{n = 1}^{{N_k}} {{p_n}}=1$.

Let $p_{N_k}=1- \sum\limits_{n = 1}^{{N_k-1}} {{p_n}}$. We have
\begin{align}
Y\left( \mathbf{p}_{k} \right) =
& \sum\limits_{n = 1}^{{N_k-1}} M_k p_n \left( 1- p_n\right)^{M_k-1} \nonumber \\
&+  M_k \left(1- \sum\limits_{n = 1}^{{N_k-1}} {{p_n}} \right) \left( \sum\limits_{n = 1}^{{N_k-1}} {{p_n}} \right)^{M_k-1}.
\end{align}
For $n=1,2,\dots,N_k-1$, take the partial derivative of $Y\left( \mathbf{p}_k \right)$ with respect to $p_n$
\begin{align}
\frac{\partial Y\left( \mathbf{p}_k \right) }{\partial p_n}=
& M_k \left(1-M_k p_n \right) \left( 1- p_n\right)^{M_k-2} \nonumber \\
&+  M_k \left(M_k -1- M_k\sum\limits_{n = 1}^{{N_k-1}} {{p_n}} \right) \left( \sum\limits_{n = 1}^{{N_k-1}} {{p_n}} \right)^{M_k-2}.
\end{align}
Let
\begin{align}
\frac{\partial Y\left( \mathbf{p}_k \right) }{\partial p_n}=0, \quad n=1,2,\dots,N_k-1.
\end{align}
The only stationary point is
\begin{align}
p_1=p_2=\dots=p_{N_k}=\frac{1}{N_k}.
\label{eq:distribution-M-le-N}
\end{align}
If $p_1=1$, then $Y\left( \mathbf{p}_k\right)=0$. Therefore, (\ref{eq:distribution-M-le-N}) achieves the global maximum. Correspondingly
\begin{align}
Y\left( \mathbf{p}_{k} \right) = {M_k} \left( 1- \frac{1}{N_k} \right)^{M_k-1},
\end{align}
which completes the proof.

\section{PROOF OF THEOREM 2}
\label{appendix:proof-theorem-max-throughput-re-rendezvous}

The detailed system setup is illustrated in Section~\ref{sec:MAC-Layer-Analysis-with-Full-SU-Information}.

It is clear that $Y^{\prime}\left( \mathbf{P}_k^{\prime} \right) = Y\left( \mathbf{p}_k \right)$ for $l_k = 0,1$. For $l_k>1$, compare $Y^{\prime}\left( \mathbf{P}_k^{\prime} \right)$ in (\ref{eq:total-throughput-in-a-slot-simplified-re-rendezvous}) and $Y\left( \mathbf{p}_k \right)$ in (\ref{eq:total-throughput-in-a-slot-simplified-max}) by
\begin{align}
 Y^{\prime}\left( \mathbf{P}_k^{\prime} \right) - Y\left( \mathbf{p}_k \right)=
& \left[ \frac{M_k N_k + l_k^2 -M_k l_k - l_k}{N_k} \right. \nonumber \\
& \left. -M_k {\left(1-\frac{1}{N_k}\right)}^{l_k}  \right]  {\left(1-\frac{1}{N_k}\right)}^{N_k-l_k-1}
\end{align}

Let
\begin{align}
f\left(l_k \right) =\frac{M_k N_k + l_k^2 -M_k l_k - l_k}{N_k} -M_k {\left(1-\frac{1}{N_k}\right)}^{l_k}.
\end{align}
The theorem can be proved by showing that $f\left(l_k \right)>0$ for $l_k>1$. It is clear that $f\left(0 \right) = f\left(1 \right) =0$. We only need to prove the first-order derivative of $f\left(l_k \right) $, $f^{\prime} \left(l_k \right)>0$. We have
\begin{align}
f^{\prime} \left(l_k \right) =\frac{2l_k -M_k -1}{N_k} -M_k {\left(1-\frac{1}{N_k}\right)}^{l_k} \ln{\left(1-\frac{1}{N_k} \right)},
\end{align}
\begin{align}
f^{\prime} \left( 1 \right) &=\frac{1-M_k}{N_k} -M_k {\left(1-\frac{1}{N_k}\right)} \ln{\left(1-\frac{1}{N_k} \right)} \nonumber \\
                            &=\frac{1 }{N_k} \left[ M_k \ln{\left(1+\frac{1}{N_k-1} \right)}^{N_k-1} - \left( M_k-1 \right)  \right].
\end{align}
We also have
\begin{align}
\frac{ M_k \ln{\left(1+\frac{1}{N_k-1} \right)}^{N_k-1}} { \left( M_k-1 \right)}
&=  \ln{\left(1+\frac{1}{N_k-1} \right)}^{\left (N_k-1 \right) \frac{M_k}{ M_k-1}} \nonumber \\
&\overset{a}{\ge}  \ln{\left(1+\frac{1}{N_k-1} \right)}^{\left (N_k-1 \right) \frac{N_k}{ N_k-1}} \label{eq:amplify-Nk} \\
&= \ln{\left(1+\frac{1}{N_k-1} \right)}^{N_k} \nonumber \\
&\overset{b}> 1.
\label{eq:inequality-to-one}
\end{align}
As a result, $f^{\prime} \left( 1 \right) >0$.
The inequality (a) results from the fact that $\frac{M_k}{ M_k-1}$ decreases as $M_k$ increases and that $M_k\le N_k$; the inequality (b) is because ${\left(1+\frac{1}{N_k-1} \right)}^{N_k}$ monotonically decreases and $ \to e$ as $N_k \to \infty$.
The second-order derivative for $f\left(l_k \right) , l_k\ge 1$ is
\begin{align}
f^{\prime \prime} \left(l_k \right)
&=   \frac{2}{N_k} -M_k {\left(1-\frac{1}{N_k}\right)}^{l_k} {\left[\ln{\left(1-\frac{1}{N_k} \right)} \right]^2} \nonumber \\
&\ge \frac{2}{N_k} -M_k {\left(1-\frac{1}{N_k}\right)} {\left[\ln{\left(1-\frac{1}{N_k} \right)} \right]^2} \nonumber \\
&=   \frac{2}{N_k} -\frac{M_k}{N_k}  \ln{\left(1+\frac{1}{N_k-1} \right)} \ln{\left(1+\frac{1}{N_k-1} \right)}^{N_k-1} \nonumber \\
&\overset{c}{>} \frac{2}{N_k}-\frac{M_k}{N_k \left(N_k-1 \right)} \label{eq:second-order-derivative-shrink}\\
&\overset{d}{\ge}0,
\end{align}
where the inequality (c) results from $\ln{\left(1+\frac{1}{N_k-1} \right)}< \frac{1}{N_k-1}$ for $N_k>1$, and because $\ln{\left(1+\frac{1}{N_k-1} \right)}^{N_k-1}$ monotonically increases and $\to e$ as $N_k \to \infty$. Inequality (d) holds because $2\le M_k \le N_k$. Therefore $f^{\prime} \left(l_k \right)>0$ for $l_k\ge 1$, and the theorem is proved.

\end{appendices}

\balance

{\small
\bibliography{IEEEabrv,mmc}
\bibliographystyle{IEEEtran}
}

\end{document}